\documentclass[11pt]{article}
\usepackage{epsfig}
\usepackage{amssymb}
\usepackage{amsmath}
\usepackage[]{times}

\makeatletter
\@addtoreset{equation}{section}
\renewcommand{\theequation} {\thesection.\arabic{equation}}
\makeatother
\newcommand{\be}{\begin{equation}}
\newcommand{\ee}{\end{equation}}
\newcommand{\bea}{\begin{eqnarray}}
\newcommand{\eea}{\end{eqnarray}}

\newcommand{\mbf}[1]{\mbox{\boldmath$#1$}}

\begin{document}

\title{Stability of Axisymmetric Liquid Bridges}
\author{Boris Rubinstein
\footnote{e-mail: bru@stowers.org}
\\ Stowers Institute for Medical Research, 1000 E 50th St, Kansas City,
MO 64110, USA}
\maketitle
\begin{abstract}
We study stability of axisymmetric liquid bridges between
two axisymmetric solid bodies in
the absence of gravity under arbitrary asymmetric perturbations
which are expanded into a set of angular
Fourier modes. We determine
the stability region boundary for every angular mode
in case of both fixed and free contact lines.
Application of this approach allows us to
demonstrate existence of stable convex
nodoid menisci between two spheres.
\end{abstract}

\section{Introduction}\label{s1}

An interface between two adjacent fluids both contacting solid(s)
is called a capillary surface, which shape depends on liquid volumes and boundary
conditions (BC) specified at the contact line where the liquids touch the
solids. A liquid bridge (LB) emerges
when a small amount of fluid (interfacing a surrounding
liquid with different properties)
contacts two (or more) solid bodies.
The LB problem has long history in both theoretical physics and pure mathematics
where the research mostly focused on two topics --
menisci shapes and related parameters (volume
$V$, surface area $A$ and surface curvature $H$) and menisci
stability.

A menisci shape study was pioneered by Delaunay \cite{Delaunay1841} who
classified all surfaces of revolution with constant mean curvature
satisfying the Young-Laplace equation (YLE). These are cylinder,
sphere, catenoid, nodoid and unduloid.
Later Beer \cite{Beer1857} found analytical solutions of YLE through
elliptic integrals and Plateau \cite{Pl1873} provided experimental
support to the LB theory. The first
explicit formulas were derived in \cite{Orr1975}
for shapes and parameters $H$, $V$ and $A$
for all meniscus types in
case of solid sphere contacting the solid plate.
A more complex case of the sphere {\it above} the plate
was considered in \cite{RubFel13}. The solutions for
meniscus shape exhibit a discrete
spectrum and are enumerated by two indices reflecting the number of inflection points
on the meniscus meridional profile and meniscus convexity. The existence of
multiple solutions \cite{RubFel13} for given volume of LB leads
to a question of menisci local stability.

The development of menisci stability theory was initiated by Sturm
\cite{Sturm1841} in appendix to \cite{Delaunay1841}, which described Delaunay's
surfaces as the solutions to an isoperimetric problem (IP).
The basis of variational theory of
stability was laid in 1870s by Weierstrass in his unpublished lectures
\cite{Weier1927} and extended by Bolza \cite{Bolz1904} and other
researchers (see Howe
\cite{Ho1887}, Knesser \cite{Knesser1900}).

The case of axisymmetric LB with fixed
contact lines (CL) was studied by Howe \cite{Ho1887} who derived
a determinant equation to produce a boundary of the
stability region under small axisymmetric perturbations.
This approach in different setups is used widely
in applications \cite{Erle1970,Gil1971}.
Forsyth \cite{Forsyth1927} considered stability of the
extremal surface of the general type under asymmetric
perturbations.
Stability of axisymmetric menisci with
free CL at solid bodies is a variational
IP with free endpoints which are allowed to run along two given planar curves
which makes a problem untractable within Howe's theory framework.

To avoid this difficulty Vogel develops
an alternative approach based on functional analysis methods.
He built an associated
Sturm-Liouville equation (SLE) for the meniscus perturbation 
with Neumann BC
instead of Dirichlet BC
for fixed CL
and established the stability criterion for LB between
parallel plates
\cite{Vogel1987}. The algorithm requires to find a solution to boundary value problem
and analyze the behavior of the two smallest eigenvalues of SLE.
Implementation of this step is extremely difficult task both both
unduloid and nodoid menisci.
This is why a single nontrivial result for catenoid meniscus
between two parallel plates
is known due to Zhou \cite{Zho97}.
The stability of LB between other solids demands an
analytical solution of boundary value problem.
Up to date this was done by Vogel only for cylindrical meniscus
between equal spheres in \cite{Vogel1999}. Another (more qualitative) result
reported in
\cite{Vogel2006} for unduloid and nodoid menisci between spheres.

A more straightforward approach was developed by
a research group headed by Myshkis (see \cite{Myshkis87} and
the references therein) which considers
a sequence of SLEs with mixed BC for the Fourier
angular modes of the perturbation.
The spectrum of $n$-th SLE
($n \ge 0$) (corresponding to $n$-th perturbation mode) consists of
discrete real values $\lambda_{n,k}, \; k \ge 1$, where
$\lambda_{n,k} < \lambda_{n,k+1}$. It was shown that
$\lambda_{n,1} < \lambda_{n+1,1},$ so that it is required only to
find sign of $\lambda^* =\min\{\lambda_{0,1},\lambda_{1,1}\}$
to establish meniscus stability. The stability boundary is given by
$\lambda^* =0$. An important development of this method is mentioned in
Sections 3.2, 3.3 in \cite{Myshkis87} for the case of asymmetric perturbations
of the axisymmetric meniscus between axisymmetric solids.

In \cite{FelRub15} and \cite{RubFel2015}
another alternative method was suggested to determine the
stability region of axisymmetric menisci with free CL
under influence of axisymmetric perturbations. It
is a development of the approach proposed in
\cite{Weier1927,Bolz1904} for the case of fixed CL.
This manuscript presents a natural extension of the method
presented in \cite{FelRub15} to the case of asymmetric perturbations.

The manuscript is organized in six sections.
In Section 2 we consider a problem of stability of axisymmetric
LB between two solids under asymmetric small perturbations
as a variational problem. We derive a general expression for
the surface energy functional with a constant liquid volume constraint
imposed on it. This expression is written explicitly for the case of
axisymmetric solid bodies; then the first and the second variations of the
functional are derived. The first variation is used to generate
YLE for the equilibrium meniscus shape and
the Dupr\'e-Young relations determining the contact angles of
the meniscus with the solids. The second variation leads to the
stability criterion of the meniscus with free CL.

In Section 3 we consider both fixed and free CL and derive the
Jacobi equation which solutions are used to establish the stability
conditions. Further following ideas of \cite{Myshkis87} we introduce
the Fourier expansion of the asymmetric perturbation into a
single axisymmetric and a set of asymmetric modes. This expansion naturally
leads to a sequence of the Jacobi equations for each perturbation mode;
then the stability conditions for each mode is derived for both
fixed and free CL.

Section 4 is devoted to computation
of the stability condition components which are used in
Section 5 to analyze the stability of unduloid and nodoid menisci
between two plates and two solid spheres.
The results are briefly discussed in Section 6.

\section{Stability problem as a variational problem}\label{s2}
Let a surface $S$ with parametrization ${\mbf \rho}(t,s)=
\{r(t,s)\cos s,r(t,s)\sin s,z(t,s)\}$,
$
0\leq s\leq 2\pi$,
is given in such a way that it is bounded by contact lines
${\bf c}_j,\ j=1,2,$
belonging to axisymmetric
solid body (SB) $S_j$ parameterized as ${\bf R}_j(\tau_j)$; the CL
itself is defined as
${\bf r}_j(t_j(s))={\bf R}_j(\tau_j(s))$.
The CL ${\bf c}_j$ is parameterized by the angular parameter $s$,
${\bf r}_j(s)={\bf R}_j(s)$ represents a curve
on the surface $S_j$,
which determines the dependencies
$t_j(s)$ and $\tau_j(s)$.
We also would need a reduced parametrization ${\bf r}(t,s)=
\{r(t,s),z(t,s)\}$ of the 
surface $S$.

Consider the first isoperimetric problem (IP--1) for a functional $E[{\mbf \rho}]$
\be
E[{\mbf \rho}]=\iint_{S}{\sf E}({\mbf \rho},{\mbf \tau},{\mbf \sigma})dt ds+
\iint_{S_1}{\sf A}_1({\bf R}_1,{\bf T}_1)d\tau_1 ds+
\iint_{S_2}{\sf A}_2({\bf R}_2,{\bf T}_2)d\tau_2 ds,
\label{e1}
\ee
with a constraint imposed on a functional $V[{\mbf \rho}]$,
\be
V[{\mbf \rho}]=\iint_{S}{\sf V}({\mbf \rho},{\mbf \tau},{\mbf \sigma})dt ds-
\iint_{S_1}{\sf B}_1({\bf R}_1,{\bf T}_1)d\tau_1 ds+
\iint_{S_2}{\sf B}_2({\bf R}_2,{\bf T}_2)d\tau_2 ds,
\label{e2}
\ee
where we
denote $f_t=\partial f/\partial t,$ and $f_{k,t}=\partial f_k/\partial t$, and
introduce two types of tangent vectors to the surface $S$:
${\mbf \tau}={\mbf \rho}_t,\;{\mbf \sigma}={\mbf \rho}_s$,
and also one to each of $S_j$: ${\bf T}_j={\bf R}_{j,\tau_j}$.
Similarly, we introduce ${\bf t}={\bf r}_t,$ and
${\bf s}={\bf r}_s,$ for the functionals $E[{\bf r}],\;V[{\bf r}]$.
The integrals over the meniscus surface $S$ and the $j$-th SB surface $S_j$
are written explicitly as
\be
\iint_{S} F dt ds =
\int_{0}^{2\pi} \!\!\!\!ds \int_{t_2(s)}^{t_1(s)} \!\!\!\!F dt,
\quad
\iint_{S_j} G_j d\tau_j ds =
\int_{0}^{2\pi} \!\!\!\!ds \int_{0}^{\tau_j(s)} \!\!\!\!G_j dt,
\label{e4a}
\ee
where $t_2(s) < t_1(s)$ for all $s$.
Denote by
$\langle{\bf a},{\bf b}\rangle$ the scalar product of two vectors ${\bf a}$
and ${\bf b},$
while the multiplication of a matrix ${\bf A}$ by a vector ${\bf b}$ is
written as ${\bf A}\cdot{\bf b}$.

Integrands ${\sf E}$ and ${\sf V}$ assumed to be
positive-homogeneous functions
of degree one in both ${\bf t}$ and ${\mbf \tau}$,
{\it e.g.},
${\sf E}({\bf r},k{\bf t},{\bf s})=k{\sf E}({\bf r},{\bf t},{\bf s})$,
resulting in identities
\bea
{\sf E}=\left\langle \frac{\partial {\sf E}}{\partial{\bf t}},{\bf t}\right\rangle
=\left\langle \frac{\partial {\sf E}}{\partial{\mbf \tau}},{\mbf \tau}\right\rangle,
\quad
{\sf V}=\left\langle \frac{\partial {\sf V}}{\partial{\bf t}},{\bf t}\right\rangle
=\left\langle \frac{\partial {\sf V}}{\partial{\mbf \tau}},{\mbf \tau}\right\rangle,
\label{e3a}
\eea
while similar relations hold for ${\sf A}_j$ and ${\sf B}_j$ w.r.t. their
argument ${\bf T}_j$:
\bea
{\sf A}_j
=\left\langle\frac{\partial {\sf A}_j}{\partial {\bf T}_j},{\bf T}_j\right\rangle,
\quad
{\sf B}_j
=\left\langle\frac{\partial {\sf B}_j}{\partial {\bf T}_j},{\bf T}_j\right\rangle.
\label{e3A}
\eea

We have to find such an extremal surface $\bar{S}$ with free CL
$\bar {\bf c}_j(s),$ located
on two given surfaces $S_j$  that the functional
$E[{\mbf \rho}]$ reaches its minimum and another functional $V[{\mbf \rho}]$
is constrained.
Define the functional $W[{\mbf \rho}]=E[{\mbf \rho}]-\lambda V[{\mbf \rho}]$
with Lagrange
multiplier $\lambda$
\be
W[{\mbf \rho}]=\!\iint_{S}\!F({\mbf \rho},{\mbf \tau},{\mbf \sigma})dtds+
\iint_{S_1}G_1({\bf R}_1,{\bf T}_1)d\tau_1 ds-
\iint_{S_2}G_2({\bf R}_2,{\bf T}_2)d\tau_2 ds,
\label{e4}
\ee
where $F={\sf E}-\lambda{\sf V}$ and $G_1=\lambda{\sf B}_1+{\sf A}_1$, $G_2=
\lambda{\sf B}_2-{\sf A}_2$. The functions $F$ and $G_j$ represent the
physical quantities of the same type ({\it e.g.}, surface area, energy, {\it etc.})
and thus have the same physical dimension.

To simplify the formulas further we use the following notation
$$
F_{\bf r} \equiv \frac{\partial F}{\partial {\bf r}},\quad
F_{\bf rt} \equiv \frac{\partial}{\partial {\bf t}}
\frac{\partial F}{\partial {\bf r}},\quad
F_{\bf tr} \equiv \frac{\partial}{\partial {\bf r}}
\frac{\partial F}{\partial {\bf t}} =
F_{\bf rt}^{T},\ etc.
$$
where ${\bf M}^T$ denotes a transposed matrix ${\bf M}$.
According to (\ref{e3a}, \ref{e3A}) we have
\be
F=\left\langle F_{\bf t},{\bf t}\right\rangle
=\left\langle F_{\mbf \tau},{\mbf \tau}\right\rangle,
\quad
G_j=
\left\langle\frac{\partial G_j}{\partial {\bf T}_j},{\bf T}_j\right\rangle.
\label{e5}
\ee
From the first relation in (\ref{e5}) we also find
\be
F_{\bf r} =
F_{\bf tr}\cdot{\bf t}
,
\quad
F_{\bf tt}\cdot{\bf t}=
{\bf 0}.
\label{e5a}
\ee
The curved meniscus surfaces are completely defined by several
differential geometry quantities:
\bea
&&
{\cal E}=\langle {\mbf \tau},{\mbf \tau} \rangle,\
{\cal G}=\langle {\mbf \sigma},{\mbf \sigma} \rangle,\
{\cal F}=\langle {\mbf \tau},{\mbf \sigma} \rangle,\
{\cal V}^2=\langle {\mbf \nu},{\mbf \nu} \rangle={\cal E}{\cal G}-{\cal F}^2,
\nonumber \\
&&
\langle {\mbf \nu},{\mbf \rho}_{tt} \rangle = {\cal VL},\quad
\langle {\mbf \nu},{\mbf \rho}_{ts} \rangle = {\cal VM},\quad
\langle {\mbf \nu},{\mbf \rho}_{ss} \rangle = {\cal VN},
\nonumber
\eea
where the cross product ${\mbf \nu}={\mbf \sigma} \times {\mbf \tau},$
defines the (unnormalized) normal vector ${\mbf \nu}$ to the surface $S$.

Before moving further we recall the standard formulas for the
computation of the surface area $A$ and the volume $V$ of the
surface defined as ${\bf r}(t,s)=
\{r_1(t,s),r_2(t,s),r_3(t,s)\}$. They read
\bea
&&A=\iint_{S} |{\mbf \nu}|\; dsdt =
\iint_{S} \sqrt{{\cal E}{\cal G}-{\cal F}^2}\; dsdt,
\quad
A_j = \iint_{S_j} |{\bf N}_j|\; ds d\tau_j,
\label{ee1}\\
&&V=\iint_{S} \langle{\mbf \nu},{\bf p}\rangle \; dsdt,
\quad
V_j=\iint_{S_j} \langle{\bf N}_j,{\bf P}_j\rangle \; ds d\tau_j,
\quad
\mbox{div}\; {\bf p} =
\mbox{div}\; {\bf P}_j = 1.
\label{ee2}
\eea
Choosing
${\bf p}=\{r_1,r_2,0\},$ and ${\bf P}_j=\{R_{j1},R_{j2},0\},$
we obtain 
\bea
V &=&\frac{1}{2}\iint_{S} \left[r_1
\left(\frac{\partial r_2}{\partial s}\frac{\partial r_3}{\partial t}-
\frac{\partial r_3}{\partial s}\frac{\partial r_2}{\partial t}
\right) -
r_2
\left(\frac{\partial r_1}{\partial s}\frac{\partial r_3}{\partial t}-
\frac{\partial r_3}{\partial s}\frac{\partial r_1}{\partial t}
\right)
\right]dsdt,
\label{ee30} \\
V_j &=&\frac{1}{2}\iint_{S_j} \left[R_{j1}
\left(\frac{\partial R_{j2}}{\partial s}\frac{\partial R_{j3}}{\partial \tau_j}-
\frac{\partial R_{j3}}{\partial s}\frac{\partial R_{j2}}{\partial \tau_j}
\right)-
R_{j2}
\left(\frac{\partial R_{j1}}{\partial s}\frac{\partial R_{j3}}{\partial \tau_j}-
\frac{\partial R_{j3}}{\partial s}\frac{\partial R_{j1}}{\partial \tau_j}
\right)
\right]ds d\tau_j.
\nonumber
\eea

We need these expressions further as the main goal of this manuscript is to
perform the stability analysis of the liquid menisci.
In this case the components ${\sf E}({\sf V})$ and ${\sf A}_j({\sf B}_j)$
of the integrands in (\ref{e4})
are proportional to the surface area (volume) of the meniscus and
two SB $S_j$, respectively:
\bea
{\sf E}&=&\gamma_{lv}\sqrt{{\cal E}{\cal G}-{\cal F}^2},
\quad
{\sf V}=\frac{1}{2}\left[r_1
\left(\frac{\partial r_2}{\partial s}\frac{\partial r_3}{\partial t}-
\frac{\partial r_3}{\partial s}\frac{\partial r_2}{\partial t}
\right) -
r_2
\left(\frac{\partial r_1}{\partial s}\frac{\partial r_3}{\partial t}-
\frac{\partial r_3}{\partial s}\frac{\partial r_1}{\partial t}
\right)
\right],
\nonumber \\
{\sf A}_j&=&(-1)^{j+1}(\gamma_{ls_j}-\gamma_{vs_j})|{\bf N}_j|,
\nonumber \\
{\sf B}_j&=&\frac{1}{2}\left[R_{j1}
\left(\frac{\partial R_{j2}}{\partial s}\frac{\partial R_{j3}}{\partial \tau_j}-
\frac{\partial R_{j3}}{\partial s}\frac{\partial R_{j2}}{\partial \tau_j}
\right)-
R_{j2}
\left(\frac{\partial R_{j1}}{\partial s}\frac{\partial R_{j3}}{\partial \tau_j}-
\frac{\partial R_{j3}}{\partial s}\frac{\partial R_{j1}}{\partial \tau_j}
\right)
\right],
\nonumber
\eea
and using these explicit expressions we find
\bea
F = \gamma_{lv}\sqrt{{\cal E}{\cal G}-{\cal F}^2} -
\lambda/2 \left[r_1
\left(\frac{\partial r_2}{\partial t}\frac{\partial r_3}{\partial s}-
\frac{\partial r_3}{\partial t}\frac{\partial r_2}{\partial s}
\right) -
r_2
\left(\frac{\partial r_1}{\partial t}\frac{\partial r_3}{\partial s}-
\frac{\partial r_3}{\partial t}\frac{\partial r_1}{\partial s}
\right)
\right].
\label{ee3b}
\eea

\subsection{Axisymmetric solid body $S_j$}
Restricting consideration to the axisymmetric SB we have
${\bf R}_j=\{R_{j}(\tau_j)\cos s,R_{j}(\tau_j)\sin s,
Z_{j}(\tau_j)\}$, where $0 \le \tau_j \le \tau_j(s),$ and find
\be
{\sf A}_j = (-1)^{j+1}(\gamma_{ls_j}-\gamma_{vs_j})R_j
\sqrt{R_j'^2+Z_j'^2},
\quad
{\sf B}_j =
R_j^2 Z_j'/2,
\label{ff1}
\ee
so that
\be
G_j = \lambda R_j^2 Z_j'/2+
(-1)^{j}(\gamma_{ls_j}-\gamma_{vs_j})R_j
\sqrt{R_j'^2+Z_j'^2}
\label{ff1a}
\ee
The SB surface area and volume read
\be
A_j = \int_{0}^{2\pi} ds
\int_{0}^{\tau_j(s)}
d\tau_j R_j\sqrt{R_j'^2+Z_j'^2},
\quad
V_j = \int_{0}^{2\pi}ds
\int_{0}^{\tau_j(s)}
d\tau_j Z_j' R_j^2/2.
\label{ff2}
\ee
Similarly, using ${\mbf \rho}(t,s)=
\{r(t,s)\cos s,r(t,s)\sin s,z(t,s)\},$ we have
$$
{\cal E}=r_t^2+z_t^2 = \langle {\bf t},{\bf t} \rangle = |{\bf t}|^2,\
{\cal G}=r^2+r_s^2+z_s^2 = r^2+\langle {\bf s},{\bf s} \rangle =
r^2+|{\bf s}|^2,\
{\cal F}=r_sr_t+z_sz_t =\langle {\bf t},{\bf s} \rangle,
$$
and obtain
\bea
{\sf E} & =& \left[
(r^2+|{\bf s}|^2)|{\bf t}|^2-
\langle {\bf t},{\bf s} \rangle^2
\right]^{1/2},
\quad
{\sf V}=r^2z_t/2,
\nonumber \\
F &=&
\gamma_{lv}\sqrt{r^2|{\bf t}|^2+|{\bf s}|^2|{\bf t}|^2-
\langle {\bf t},{\bf s} \rangle^2}-\frac{\lambda r^2z_t}{2}.
\label{ff3}
\eea
If the surface $S$ is axisymmetric too the contact lines
transform into circles, and its surface area and volume
read
\be
A = 2\pi
\int_{t_2}^{t_1}
dt \;r\sqrt{r_t^2+z_t^2},
\quad
V= \pi
\int_{t_2}^{t_1}
dt \;z_t r^2,
\label{ff4}
\ee
so that (\ref{ff3}) reduces to
\be
F = \gamma_{lv}\sqrt{r^2(r_t^2+z_t^2)}-\frac{\lambda r^2z_t}{2}.
\label{ff5}
\ee
The variational problem with (\ref{ff5}) and (\ref{ff1a}) under axisymmetric
perturbations was considered in \cite{FelRub15}.
It should be underscored here that the selection of axisymmetric contact surfaces $S_j$
{\it does not} imply that the surface $S$ should be axisymmetric too.

The goal of this manuscript is to develop a framework for the description of the
stability of asymmetric meniscus under general {\it asymmetric} small perturbations.
This requires a consideration of the functional $W$ with $F$ and $G_j$ given by
(\ref{ff3}) and (\ref{ff1a}), respectively. We impose only one restriction on this
setup, namely, we require that the contact lines with the axisymmetric solid bodies
should be {\it circular}. Then the
integration of $F$ should be performed in the following range of $t$ values
$t_2 \le t \le t_1$, where both limits are independent of $s$. Correspondingly,
the upper integration limit $\tau_j$ for $G_j$ also does not depend on $s$.


\subsection{Meniscus surface perturbation}
Introduce a six-dimensional vector
${\bf p}(t,s)=\{r,z,r_t,z_t,r_s,z_s\}
\equiv \{{\bf r},{\bf t},{\bf s}\},$ and
calculate total variation of the functional, ${\mathbb D}W={\mathbb D}_0W+
{\mathbb D}_1W-{\mathbb D}_2W$, where each term
represents the variation of the corresponding term of $W[{\bf r}]$ in (\ref{e4}).
Consider the first term,
denoting a small variation of the surface $S$ as
${\bf u}(t,s)=\{u(t,s),v(t,s)\},$
restricted by a
condition on CL that it {\it should always} belong to the surface $S_j$:
$$
{\bf r}(t_j)+{\bf u}(t_j(s),s) =
{\bf R}_j(\tau_j+\delta\tau_j(s)),
$$
so that we arrive at the expansion
$$
{\bf u}(t_j(s),s) =
\sum_{k=1}^{\infty}{\bf u}_k(\tau_j(s),s),
\quad
{\bf u}_k(t_j(s),s) =
\frac{1}{k!}
\frac{d^k{\bf R}_{j}
}{d\tau_j^k}\;
\delta^k\tau_j(s).
$$
Thus we obtain in the lowest orders
\be
{\bf u}_1(t_j(s),s)=\frac{d{\bf R}_j}{d\tau_j}\;\delta\tau_j(s) =
{\bf T}_j\delta\tau_j,\quad
{\bf u}_2(t_j(s),s)=\frac{1}{2}\frac{d{\bf T}_j}{d\tau_j}\;\delta^2\tau_j(s).
\label{ea13}
\ee
The variation due to integrand perturbation is found as
\bea
{\mathbb D}_0W&=&\int_{0}^{2\pi}ds
\int_{t_2}^{t_1}[\Delta_1 F+
\Delta_2 F+\ldots]\; dt,
\label{e6}\\
\Delta_1F&=&\langle F_{\bf p},{\bf h}\rangle,
\label{e7}\\
\Delta_2F&=&\frac{1}{2}
\langle{\bf h},F_{\bf pp}\cdot{\bf h}\rangle,
\label{e8}
\eea
where
${\bf h}=\{u,v,u_t,v_t,u_s,v_s\}
\equiv \{{\bf u},{\bf u}_t,{\bf u}_s\}$.
The variation ${\mathbb D}_jW$ due to perturbation of the $j$-th CL
parameterized by $\delta \tau_j(s)$ reads
\bea
{\mathbb D}_jW=
\int_{0}^{2\pi}\!\!ds
\int_0^{\tau_j+\delta\tau_j(s)}
\!\!\!\!\!\!\!\!\!\!\!\!\!\!\!\!G_j\;d\tau_j-
\int_{0}^{2\pi}\!\!ds
\int_0^{\tau_j}\!\!\!\!\!G_j\;d\tau_j =
\int_{0}^{2\pi}\!\!ds
\int_{\tau_j}^{\tau_j+\delta\tau_j(s)}
\!\!\!\!\!\!\!\!\!\!\!\!\!\!\!\!\!\!\!G_j\;d\tau_j\;.
\label{e6a}
\eea
Further we need the inner integral in (\ref{e6a})
expanded up to the terms quadratic in
$\delta\tau_j$:
\be
\int_{\tau_j}^{\tau_j+\delta\tau_j(s)}
\!\!\!\!\!\!\!\!\!\!\!\!\!G_j\;d\tau_j=
G_j^*\delta\tau_j(s)+
\frac{1}{2}
\frac{dG_j^*}{d\tau_j}
\;[\delta\tau_j(s)]^2+\ldots,
\quad
G_j^*=
G_j(\tau_j).
\label{e6b}
\ee
Using this expansion we find
\bea
{\mathbb D}_jW=
\int_{0}^{2\pi}
\left[G_j^*\delta\tau_j(s)+
\frac{1}{2}
\frac{dG_j^*}{d\tau_j}
\;[\delta\tau_j(s)]^2+\ldots\right]ds.
\label{e6c}
\eea

\subsection{First Variation $\delta W$}\label{s21}
\begin{figure}[h!]\begin{center}
\psfig{figure=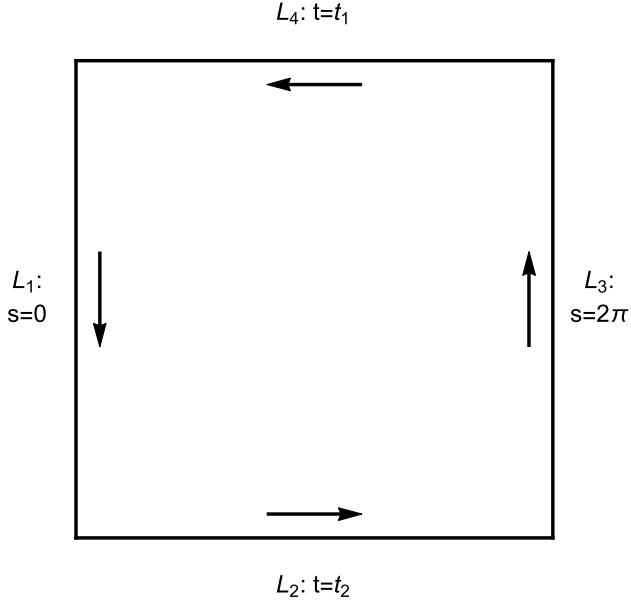,height=8cm}
\end{center}
\caption{Sketch of the integration contour in the
$\{s,t\}$ coordinates in case of axisymmetric solid bodies and
circular contact lines $t=t_1$ and $t=t_2$.}
\label{fig01}
\end{figure}

Using expressions (\ref{e6}) for ${\mathbb D}_0W$ and ${\mathbb D}_jW$
of the terms linear in $\delta\tau_j$ and ${\bf h}$, calculate $\delta W$
\bea
\delta W=
\int_{0}^{2\pi}ds\left[
\int_{t_2}^{t_1}dt \Delta_1 F+
G_1^*\delta\tau_1(s)-
G_2^*\delta\tau_2(s)
\right].
\label{e11}
\eea
The explicit expression for the integrand variation reads:
$$
\Delta_1 F =
\langle F_{\bf r},{\bf u}\rangle +
\langle F_{\bf t},{\bf u}_t\rangle +
\langle F_{\bf s},{\bf u}_s\rangle,
$$
Following \cite{Forsyth1927} integrate the relations
$$
\frac{\partial}{\partial t} \langle F_{\bf t},{\bf u}\rangle =
\langle F_{\bf t},{\bf u}_t\rangle +
\langle \partial F_{\bf t}/\partial t,{\bf u}\rangle,
\quad
\frac{\partial}{\partial s} \langle F_{\bf s},{\bf u}\rangle =
\langle F_{\bf s},{\bf u}_s\rangle +
\langle \partial F_{\bf s}/\partial s,{\bf u}\rangle,
$$
and use the Green's theorem
$$
\iint_{S} dsdt\left(\frac{\partial Q}{\partial t}-\frac{\partial P}{\partial s}\right)
=\int_{L} (P dt + Q ds),
$$
to find the first term in (\ref{e11})
\be
\iint_{S}dsdt \Delta_1 F  =
\iint_{S}dsdt
\langle {\bf \mbf{\delta} F},{\bf u}\rangle
+\int_{L} ds \langle F_{\bf t},{\bf u}\rangle
-\int_{L} dt \langle F_{\bf s},{\bf u}\rangle,
\quad
{\bf \mbf{\delta} F} = F_{\bf r}
-\frac{\partial F_{\bf t}}{\partial t}
-\frac{\partial F_{\bf s}}{\partial s},
\label{e12a}
\ee
where $L$ in the last two integrals denotes the boundary
of the integration region. Consider computation of these integrals
in an important particular case of the {\it axisymmetric} surfaces $S_j$
using the cylindrical coordinates and assuming without loss of generality that the
variable $s$ denotes the polar angle ($s_2=0 \le s \le s_1=2\pi$),
while $t$ covers the range $t_2 \le t \le t_1$,
The integration contour $L$ consists of four segments $L_k$ shown in
Figure \ref{fig01}: $L_1:\{s=0, t_2\le t\le t_1\},\
L_2:\{0\le s\le 2\pi, t=t_2\},\
L_3:\{s=2\pi, t_2\le t\le t_1\},\
L_4:\{0\le s\le 2\pi, t=t_1\}$.
The integration results w.r.t. $t$ along
the lines $s=0$ and $s=2\pi$ cancel each other and thus
we have to find the contributions for $L_2$ and $L_4$ only.
As the integration along these lines goes in opposite directions
we have for the contour integral over $s$
\be
\int_{L} ds \langle F_{\bf t},{\bf u}\rangle =
\int_{0}^{2\pi}\!\!\! ds \left[\langle F_{\bf t},{\bf u}\rangle|_{t=t_1}-
\langle F_{\bf t},{\bf u}\rangle|_{t=t_2}\right].
\label{e13a}
\ee
Finally, the expression (\ref{e12a}) reduces to
\be
\iint_{S}dsdt \Delta_1 F  =
\iint_{S}dsdt
\langle {\bf \mbf{\delta} F},{\bf u}\rangle
+\int_{0}^{2\pi}\!\!\! ds \left[\langle F_{\bf t},{\bf u}\rangle|_{t=t_1}-
\langle F_{\bf t},{\bf u}\rangle|_{t=t_2}\right],
\label{e14}
\ee
and we write
\be
\delta W=
\iint dtds \langle {\bf \mbf{\delta} F},{\bf u}\rangle+
\int_{0}^{2\pi}\!\!\!\!\!\!ds\left[
G_1^*\delta\tau_1(s)+
\langle F_{\bf t},{\bf u}\rangle|_{t=t_1}-
G_2^*\delta\tau_2(s)-
\langle F_{\bf t},{\bf u}\rangle|_{t=t_2}
\right],
\label{e14a}
\ee
where the terms in (\ref{e14}) are paired with the
boundary terms in (\ref{e11}), while the double integral
should vanish to guarantee vanishing of the first variation.
As the small perturbation ${\bf u}$ is arbitrary we conclude that
the following condition should hold:
\be
{\bf \mbf{\delta} F} = F_{\bf r}
-\frac{\partial F_{\bf t}}{\partial t}
-\frac{\partial F_{\bf s}}{\partial s} = {\bf 0},
\label{e13b}
\ee
which corresponds to two Euler-Lagrange (EL) equations.
The EL equations (\ref{e13b}) determine a surface of an asymmetric meniscus
with {\it circular} CL on both axisymmetric SB.
Search of general solutions of (\ref{e13b}) represents a difficult problem,
and it is out of scope of this manuscript.

We further
restrict ourself to the case of {\it axisymmetric} menisci
as liquid bridge equilibrium surface, and
thus we simplify equations (\ref{e13b}) into
\be
F_{\bf r}-\frac{d F_{\bf t}}{dt}= {\bf 0},
\label{e13c}
\ee
assuming the solution ${\bf \bar r}={\bf \bar r}(t)$.
Setting $\lambda = 2\gamma_{lv} H$, where $H$ is the mean curvature,
we obtain from (\ref{e13c}):
$$
r_{tt} = - z_t(2H - z_t/r),
\quad
z_{tt} = r_t(2H - z_t/r),
$$
from which it follows that a condition
$r_t^2+z_t^2=\langle {\bf t},{\bf t}\rangle=1,$ holds.
The definition of $\lambda$ should be used in (\ref{ff3}) which
after rescaling to $\gamma_{lv}$ takes
two equivalent forms which will be used further on
\be
F=\sqrt{r^2|{\bf t}|^2+|{\bf s}|^2|{\bf t}|^2-
\langle {\bf t},{\bf s} \rangle^2}-H r^2z_t=
\sqrt{r^2|{\bf t}|^2+
\langle {\bf n},{\bf s} \rangle^2}-H r^2z_t.
\label{e13d}
\ee

In (\ref{e14a}) we retain only the terms
linear in $\delta\tau_j$, {\it i.e.},
proportional to ${\bf u}_1$; the higher order terms will contribute to
the second and higher variations.
Using (\ref{ea13}) we find that the first variation vanishes when
(\ref{e13b}) holds along with
\bea
0&=&
\int_{0}^{2\pi}\!\!\!\!\!\!ds
\left[
G_1^*\delta\tau_1(s)+
\langle F_{\bf t},{\bf u}_1\rangle|_{t=t_1}-
G_2^*\delta\tau_2(s)-
\langle F_{\bf t},{\bf u}_1\rangle|_{t=t_2}
\right] \nonumber \\
&=&
\int_{0}^{2\pi}\!\!\!\!\!\!ds
\left[
G_1^*+
\langle F_{\bf t}(t_1),{\bf T}_1\rangle
\right]\delta\tau_1(s)-
\int_{0}^{2\pi}\!\!\!\!\!\!ds\left[
G_2^*+
\langle F_{\bf t}(t_2),{\bf T}_2\rangle
\right]\delta\tau_2(s).
\label{e14b}
\eea
Due to arbitrariness of the CL perturbation $\delta\tau_j(s)$ we conclude that
two boundary conditions should hold
\be
G_j^*+
\langle F_{\bf t}(t_j),{\bf T}_j\rangle=0.
\label{e15b}
\ee
The transversality conditions
(\ref{e15b}) are known as the Dupr\'e-Young relations for the contact angle $\theta_j$
of the meniscus with the $j$-th SB,
\bea
\frac{\gamma_{ls_j}-\gamma_{vs_j}}{\gamma_{lv}}+\cos\theta_j=0,
\quad
\cos\theta_j= (-1)^{j+1}
\frac{\langle{\bf t}_j,{\bf T}_j\rangle}{|{\bf t}_j||{\bf T}_j|} =
(-1)^{j+1}
\frac{\langle{\bf n}_j,{\bf N}_j\rangle}{|{\bf n}_j||{\bf N}_j|},
\label{h4}
\eea
where ${\bf n}$ denotes the normal to the meridional cross section of the meniscus, {\it i.e.},
$\left\langle{\bf t},{\bf n}\right\rangle=0$.

Introduce a projection $W$ of the perturbation ${\bf u}$ on the
normal ${\mbf \nu}$ to the meniscus:
$W(t,s) = \left\langle{\bf u},{\mbf \nu}\right\rangle$.
At the endpoints $t_j$
this quantity does not depend on $s$ and $W(t)$ has the values
depending on $\delta\tau_j$,
\be
W(t_j)=R_j(\tau_j^*)\eta(t_j,\tau_j^*)\delta\tau_j+\ldots,\quad
\eta(t_j,\tau_j^*)=\eta_j=
\left\langle{\bf T}_j,{\bf n}(t_j)\right\rangle.
\label{e15}
\ee
Comparison of (\ref{h4}) with (\ref{e15}) implies that
$\eta_j$ is proportional to $\sin \theta_j$.
Further we use a projection $w$ of the perturbation ${\bf u}$ on the
normal ${\bf n}$: $w(t,s)=\left\langle{\bf u},{\bf n}\right\rangle$,
so that $W(t_j)=R_j(\tau_j^*)w(t_j)$.


The solution ${\bf r}={\bf\bar r}(t)$ of
(\ref{e13c}) together with (\ref{e15b})
provides the extremal value of
$E[{\bf r}]$ constrained by $V[{\bf r}]=1$.
This extremal curve cannot intersect any of the solid bodies,
except the contact at the points $t_j$.
It can be satisfied when a simple geometric condition on the tangents to the
extremal curve and the solid at the contact point holds.
This existence condition
can be expressed as $\eta_j \ge 0$, and $\eta_j=0$
defines a boundary of a meniscus {\it existence region}.

\subsection{Second Variation $\delta^2W$}\label{s22}
Use in (\ref{e6}) the terms quadratic in $\delta\tau_j$ and ${\bf h}$,
and calculate the second variation $\delta^2W$,
\be
\delta^2W=\int_{0}^{2\pi}\!\!\!\!\!\!\!ds
\left[
\int_{t_2}^{t_1}\!\!\!\!\!\!\!\Delta_2 F dt
+\left\langle F_{\bf t},{\bf u}_2(t)\right\rangle_{t_2}^{t_1}
+\frac{1}{2}\left(
\frac{dG_1}{d\tau_1}[\delta\tau_1(s)]^2-\frac{dG_2}{d\tau_2}[\delta\tau_2(s)]^2\right)
\right]
=\int_{0}^{2\pi}\!\!\!\!\!\!\!ds \delta^2 \tilde W(s) ,
\label{e20}
\ee
Here the term $\left\langle F_{\bf t},{\bf u}_2(t)\right\rangle$ is added
due to the reason described above in discussion of (\ref{e14a}).
Substituting ${\bf u}_2(t)$ from (\ref{ea13}) into the
last expression we obtain for the inner integral in (\ref{e20})
\bea
\delta^2 \tilde W(s)&=&
\int_{t_2}^{t_1}\Delta_2 Fdt+
\frac{1}{2}\left(\left\langle F_{\bf t}(t_1) 
,\frac{d{\bf T}_1}{d\tau_1}\right\rangle
+\frac{dG_1}{d\tau_1}\right)[\delta\tau_1(s)]^2
\nonumber \\
&-&
\frac{1}{2}\left(\left\langle F_{\bf t}(t_2) 
,\frac{d{\bf T}_2}{d\tau_2}\right\rangle
+\frac{dG_2}{d\tau_2}\right)[\delta\tau_2(s)]^2.
\label{e20aa}
\eea
First compute the general expression for $\Delta_2 F$:
\bea
\Delta_2 F & = &
\frac{1}{2}\left\langle {\bf u},F_{\bf rr}\cdot{\bf u}\right\rangle
+\left\langle {\bf u},F_{\bf tr}\cdot{\bf u}_{t}\right\rangle
+\frac{1}{2}\left\langle {\bf u}_{t},F_{\bf tt}\cdot{\bf u}_{t}\right\rangle
\nonumber \\
&+&\frac{1}{2}\left\langle {\bf u}_{s},F_{\bf ss}\cdot{\bf u}_{s}\right\rangle
+\left\langle {\bf u},F_{\bf sr}\cdot{\bf u}_{s}\right\rangle
+\left\langle {\bf u}_{s},F_{\bf ts}\cdot{\bf u}_{t}\right\rangle.
\nonumber
\eea
Recalling that the meniscus equilibrium
axisymmetric surface ${\bf\bar r}(t)$ depends only on $t$, we can check by
direct computation that last two terms in the above expression vanish,
and we end up with
\be
\Delta_2 F =
\frac{1}{2}\left\langle {\bf u},F_{\bf rr}\cdot{\bf u}\right\rangle
+\left\langle {\bf u},F_{\bf tr}\cdot{\bf u}_{t}\right\rangle
+\frac{1}{2}\left\langle {\bf u}_{t},F_{\bf tt}\cdot{\bf u}_{t}\right\rangle
+\frac{1}{2}\left\langle {\bf u}_{s},F_{\bf ss}\cdot{\bf u}_{s}\right\rangle.
\label{e20a}
\ee
Denote $\delta^2_B \tilde W=\int_{t_2}^{t_1}\Delta_2 Fdt$
and generalizing an approach of Weierstrass
\cite{Weier1927}, pp.132-134 (see also Bolza \cite{Bolz1904}, p.206) represent
it in terms of small perturbation ${\bf u}_1$
and $w(t,s)$
\bea
&&
\delta^2_B \tilde W=
\frac{1}{2}
\left[
\Xi_0[w]+
\left\langle {\bf u}_1,{\bf {\widehat L}}\cdot{\bf u}_1
\right\rangle_{t_2}^{t_1}
\right],
\quad
{\bf {\widehat L}} =
F_{\bf tr}
-H_1(t)\;{\bf n}'\otimes{\bf n},
\label{e21}  \\
&&
\Xi_0[w]=\int_{t_2}^{t_1}
\!\!\left[H_1(t)w_t^2(t,s)+H_4(t)w_s^2(t,s)+H_2(t)w^2(t,s)\right]dt,
\label{e22}
\eea
where $H_1(t),\;H_2(t),$ and $H_4(t)$ are defined through matrix relations
\be
F_{\bf tt}
=H_1(t)\;{\bf n}\otimes{\bf n},\quad
F_{\bf ss}
=H_4(t)\;{\bf n}\otimes{\bf n},\quad
F_{\bf rr}
-\frac{\partial{\bf {\widehat L}}}{\partial t}-
H_1(t)\;{\bf n}'\otimes{\bf n}'=
H_2(t)\;{\bf n}\otimes{\bf n},
\label{e23}
\ee
$\otimes$ denotes the outer product of two vectors,
${\bf n}'=d{\bf n}/dt,$
and ${\bf n}(t)$ denotes the normal to the
meridional cross section of the meniscus ${\bf \bar r}(t)$.
The expression (\ref{e22}) for $\Xi_0[w]$ generalizes formula (2.17)
in \cite{FelRub15} to the case of
asymmetric perturbations.
The relation (\ref{e20aa}) reads
\bea
&&\delta^2\tilde W=\delta^2_B\tilde W+
\xi_1[\delta\tau_1(s)]^2-\xi_2[\delta\tau_2(s)]^2,
\label{e24}
\\
&&\xi_j=\frac{1}{2}
\left(
\left\langle
F_{\bf t}(t_j), 
\frac{d{\bf T}_j}{d\tau_j}
\right\rangle+
\left\langle
\frac{\partial G_j}{\partial {\bf R}_j},
{\bf T}_j
\right\rangle+
\left\langle
\frac{\partial G_j}{\partial {\bf T}_j},
\frac{d{\bf T}_j}{d\tau_j}
\right\rangle
\right).
\label{e25}
\eea
Substitute ${\bf u}_1(t_j)$ from (\ref{ea13}) into (\ref{e21}) and
combine it with (\ref{e24}) to find 
\be
\delta^2W=\int_{0}^{2\pi}\!\!\!\!\!ds
\left[
\frac{1}{2}\Xi_0[w]+K_1[\delta\tau_1(s)]^2-K_2[\delta\tau_2(s)]^2
\right],\quad
K_j=\xi_j+\frac{1}{2}
\left\langle {\bf T}_j,{\bf {\widehat L}}(t_j)\cdot{\bf T}_j\right\rangle.
\label{e26}
\ee
Using the definition (\ref{e21}) compute the following term in the above expression
$$
\left\langle {\bf T}_j,{\bf {\widehat L}}(t_j)\cdot{\bf T}_j\right\rangle =
\left\langle {\bf T}_j,{\bf F_{tr}}(t_j)\cdot{\bf T}_j\right\rangle -
H_1(t_j)\langle {\bf n}_j',{\bf T}_j\rangle
\langle {\bf n}_j,{\bf T}_j\rangle.
$$
Introducing $\eta_j' = \langle {\bf n}_j',{\bf T}_j\rangle,$ we find
\be
K_j=\frac{1}{2}
\left(\left\langle
F_{\bf t}(t_j)+\frac{\partial G_j}{\partial {\bf T}_j},
\frac{d{\bf T}_j}{d\tau_j}
\right\rangle+
\left\langle
F_{\bf tr}(t_j)\cdot{\bf T}_j+\frac{\partial G_j}{\partial {\bf R}_j},
{\bf T}_j
\right\rangle
-H_1(t_j)\eta_j\eta_j'
\right).
\label{e31}
\ee
Multiply ${\bf {\widehat L}}(t)$ by the vector ${\bf t}$;
using the relation (\ref{e5a}) and $\langle {\bf n}, {\bf t} \rangle =0,$
from (\ref{e22}) we obtain (see also \cite{Bolz1909}, p. 226):
\be
{\bf {\widehat L}}(t)\cdot{\bf t} = F_{\bf tr}\cdot{\bf t}  = F_{\bf r},
\label{e29}
\ee
Show that the EL equations (\ref{e13c}) imply the following symmetry:
${\bf {\widehat L}} = {\bf {\widehat L}}^T$. To this end rewrite (\ref{e13c}) performing
the differentiation w.r.t. $t$ explicitly and use (\ref{e23}):
$$
F_{\bf r}
-\frac{\partial F_{\bf t}}{\partial t}
=F_{\bf r} -F_{\bf rt} \cdot{\bf t}-
F_{\bf tt}\cdot{\bf t}'=
F_{\bf r}
-{\bf {\widehat L}}^T \cdot{\bf t} -
H_1(t)\left\langle {\bf n}',{\bf t}\right\rangle {\bf n}-
H_1(t)\left\langle {\bf n},{\bf t}'\right\rangle {\bf n}.
$$
Noting that $\left\langle {\bf n}',{\bf t}\right\rangle +
\left\langle {\bf n},{\bf t}'\right\rangle = \left\langle {\bf n},{\bf t}\right\rangle' = 0$,
we find
$$
F_{\bf r} 
-\frac{\partial F_{\bf t}}{\partial t}=
F_{\bf r}-{\bf {\widehat L}}^T\cdot {\bf t} = {\bf 0},
$$
and recalling (\ref{e29}) we arrive at
$({\bf {\widehat L}}-{\bf {\widehat L}}^T)\cdot{\bf t}={\bf 0}$.
We obtain
\be
F_{\bf r}
-\frac{\partial F_{\bf t}}{\partial t}
=({\bf {\widehat L}}-{\bf {\widehat L}}^T)\cdot{\bf t} =
T {\bf n},
\
\mbox{where}
\ \
T=L_{12}-L_{21}=F_{rz_t}-F_{zr_t}+H_1
\left\langle {\bf n}',{\bf t}\right\rangle = 0.
\label{e32}
\ee
Thus the EL equations (\ref{e13c}) are equivalent to
single Young-Laplace equation (\ref{e32}).
The computation of the first variation $\delta V$ is done similarly
(\cite{Bolz1904}, p.215) and it produces
\be
\delta V=2\int_{0}^{2\pi}\!\!\!\!\!ds
\int_{t_2}^{t_1} \!\!\!\!H_3(t)w(t,s)dt = 0,
\label{e19}
\ee
where $H_3$ is determined through the relations
\be
{\sf V}_{\bf r}
-\frac{d{\sf V}_{\bf t}}{dt} =
H_3(t) {\bf n},
\
H_3(t) = {\sf V}_{rz_t}-{\sf V}_{zr_t}+{\sf V}_1
\left\langle {\bf n}',{\bf t}\right\rangle,
\
V_{\bf tt}
={\sf V}_1\;{\bf n}\otimes{\bf n}.
\label{e19a}
\ee
Consider the second expression in (\ref{e23}) determining the function $H_2$.
Using the definition (\ref{e22}) of the matrix ${\bf {\widehat L}}$ we
have
$$
H_2(t)\;{\bf n}\otimes{\bf n}=
F_{\bf rr}-\frac{\partial {\bf {\widehat L}}}{\partial t}-
H_1(t)\;{\bf n}'\otimes{\bf n}'=
F_{\bf rr}-\frac{\partial }{\partial t}F_{\bf tr}
+\left(H_1(t)\;{\bf n}'\right)'\otimes{\bf n}.
$$
Using (\ref{e32}) we have,
\be
F_{\bf rr}-\frac{\partial }{\partial t}F_{\bf tr} =
\frac{\partial}{\partial{\bf r}}
\left(F_{\bf r}-\frac{\partial F_{\bf t}}{\partial t}\right)=
T_{\bf r}\otimes {\bf n},
\quad
H_2\;{\bf n} = T_{\bf r}+(H_1\; {\bf n}')',
\label{e33}
\ee
and find
\be
z_t H_2 = \frac{\partial (H_1 z_{tt})}{\partial t} +\frac{\partial T}{\partial r},
\quad
r_t H_2 =\frac{\partial (H_1 r_{tt})}{\partial t}-\frac{\partial T}{\partial z}.
\label{e34}
\ee
Using the definition (\ref{e32}) rewrite the above relations
\be
z_t H_2 = \frac{\partial (H_1 z_{tt})}{\partial t}+(F_{rrz_t}-F_{rzr_t}),
\quad
r_t H_2 = \frac{\partial (H_1 r_{tt})}{\partial t}-(F_{rzz_t}-F_{zzr_t}).
\label{e36}
\ee
The explicit expression for the functions $H_i(t)$ for the integrand
$F$ in (\ref{ff3}) read
\be
H_1 = H_3 = r,
\quad
H_2 = (r r'')'/r',
\quad
H_4 = 1/r.
\label{e37}
\ee

\section{Boundary conditions}
\label{s3}
To study stability of extremal curve ${\bf\bar r}(t)$ w.r.t. small
perturbations it is convenient to consider two cases which differ
by the conditions imposed on the perturbed meniscus CL -- fixed CL and free CL.

\subsection{Fixed contact lines}
\label{s31}

The first case is when ${\bf\bar r}(t)$ is perturbed in the interval
$(t_2,t_1),$ but the CLs are fixed,
\be
{\bf u}(t_j)={\bf 0}, \quad w(t_j)=0,\quad j=1,2.
\label{f1}
\ee
Start with the second isoperimetric problem (IP--2) associated with extremal
perturbations ${\bf u}(t)$ in vicinity of ${\bf\bar r}(t)$ with BC
(\ref{f1}) and constraint of the volume conservation (\ref{e19})
\bea
\Xi_1[w]=\int_{0}^{2\pi}\!\!\!\!\!ds
\int_{t_2}^{t_1}H_3(t)w(t,s)dt=0,
\label{f2}
\eea
involving the perturbation $w(t)$. Substituting (\ref{f1}) into (\ref{e21}) we
arrive at the classical isoperimeteric problem with the second variation $\Xi_0[w]$.
Analyzing the problem with functional $\Xi_2[w]=\Xi_0[w]+2\mu\Xi_1[w],$
where $\mu$ denotes a Lagrange multiplier,
\bea
\Xi_2[w]=\int_{0}^{2\pi}\!\!\!\!\!ds
\int_{t_2}^{t_1}\!\!\!\!\!dt[
H_1(t)w_t^2+H_4(t)w_s^2+H_2(t)w^2+2\mu H_3(t)w
],
\label{f3}
\eea
write the EL equation with BC (\ref{f1}) for
extremals $w(t,s)$ which is the inhomogeneous Jacobi equation
\be
(H_1w_t)_t+H_4 w_{ss}-H_2 w=\mu H_3,
\label{f4}
\ee
with the boundary conditions $w(t_1,s)=w(t_2,s)=0$.

\subsection{Free contact lines}
\label{s32}

Consider a case when ${\bf\bar r}(t)$ is perturbed at interval
$[t_2,t_1]$ including both CL.
The nonintegral term in (\ref{e26}) is fixed and in general case it does not
vanish.
Following ideology of stability theory we have to find conditions when $\delta^2W$ is
positive definite in vicinity of extremal curve constrained by (\ref{e2}). Since
the only varying part in (\ref{e26}) is the functional $\Xi_0[w]$, this brings
us to IP--2 with one indeterminate function $w(t,s)$: find the extremal $\bar{w}
(t,s)$ providing $\Xi_0[w]$ to be positive definite in vicinity of $\bar{w}(t)$
and preserving $\Xi_1[w]$.


Using the reasoning presented in \cite{FelRub15}
write $w(t,s)$ in vicinity of extremal perturbation $\bar{w}(t)$ as follows,
\bea
w(t,s)&=&\bar{w}(t,s)+\varepsilon(t,s),\quad\varepsilon(t_1,s)=\varepsilon(t_2,s)=0,
\quad\varepsilon(t,0)=\varepsilon(t,2\pi),
\nonumber \\
\Xi_1[\varepsilon]&=&\int_{0}^{2\pi}ds
\int_{t_2}^{t_1}dt H_3\varepsilon(t,s)=0,
\label{g3}
\eea
where a perturbation $\varepsilon(t)$ does not break BC (\ref{e15}),
and preserves the volume conservation condition (\ref{f2}).

Find the first and second
variations of functional $\Xi_2[w]$ defined in (\ref{f3}),
\bea
\delta\Xi_2[w]\!&=&\!2\int_{0}^{2\pi}ds
\int_{t_2}^{t_1}\left[-(H_1\bar{w}_t)_t-H_4\bar{w}_{ss}+H_2\bar{w}+\mu
H_3\right]\varepsilon(t)\;dt,\label{g4}\\
\delta^2\Xi_2[w]\!&=&\!\int_{0}^{2\pi}ds
\int_{t_2}^{t_1}\left[H_1\varepsilon_t^2+H_4\varepsilon_s^2+
H_2\varepsilon^2\right]dt,\label{g5}
\eea
The first variation $\delta\Xi_2[w]$ vanishes at the extremal $\bar{w}(t)$
satisfying the inhomogeneous Jacobi equation (\ref{f4}). Regarding the
second
variation $\delta^2\Xi_2[w]$ it completely coincides with $\Xi_0[w]$, as well as
BC and volume constraint (\ref{g3}) are coinciding with similar BC (\ref{f1})
and constraint (\ref{f2}) in the isoperimetric problem with fixed endpoints
(Section \ref{s31}).
%

\subsection{Fourier expansion}
\label{s33}

Consider a homogeneous version of (\ref{f4})
\be
(H_1w_t)_t+H_4 w_{ss}-H_2 w=0,
\label{j01}
\ee
and seek one of its fundamental solutions using the separation of variables
$w(t,s)=T(t)S(s)$. Substituting this ansatz into (\ref{j01}) we obtain
$S(H_1T')'+H_4 TS''-H_2 TS=0,$ leading to
\be
[(H_1T')'/T-H_2]/H_4=-S''/S=n^2,
\label{j02}
\ee
where $n^2$ is the separation constant. These two equations can be written as
\be
S''+n^2S = 0,
\quad
(H_1T')'-H_2 T-n^2 H_4 T=0,
\label{j03}
\ee
where the first equation naturally leads to Fourier angular modes
$S_n(s) = S_0 \exp(ins),$ for integer $n$.

Following \cite{Myshkis87} expand the perturbation ${\bf u}(t,s)$
and its components ${\bf u}_k(t,s)$
into Fourier series in the angular variable
$s$ as follows:
\be
{\bf u}_k(t,s) =
{\bf u}_k^{(0)}(t) +
\sum_{n=1}^{\infty}
\left[{\bf u}_k^{(n)}(t) \exp(ins)+c.c.\right],
\label{k01}
\ee
where the term ${\bf u}_k^{(0)}(t)$ describes
axisymmetric perturbation, while the remaining terms are responsible for the
asymmetric perturbations; $c.c.$ stands for complex conjugate.
Similarly, we write
\be
w(t,s) =
w^{(0)}(t) +
\sum_{n=1}^{\infty}
[w^{(n)}(t) \exp(ins)+c.c.].
\label{k02}
\ee
The perturbation of the $j$-th CL described by the function $\delta\tau_j(s)$ is also
expanded
\be
\delta\tau_j(s) =
\delta\tau_j^{(0)} +
\sum_{n=1}^{\infty}
[\delta\tau_j^{(n)} \exp(ins)+c.c.].
\label{k03}
\ee
The complex Fourier amplitudes $\delta\tau_j^{(n)}$ are
computed through inverse complex Fourier transform.
Substitution of (\ref{k02})
into (\ref{e19}) produces a series of the conditions
$$
\delta V_n=2\int_{0}^{2\pi}\!\!\!\!\!\exp(ins) ds
\int_{t_2}^{t_1} \!\!\!\!H_3(t)w^{(n)}(t)dt = 0,
$$
which lead to a single nontrivial condition for the axisymmetric mode
\be
\int_{t_2}^{t_1} \!\!\!\!H_3(t)w^{(0)}(t)dt = 0,
\label{k06}
\ee
while for the asymmetric modes ($n \ge 1$)
the corresponding conditions are satisfied
identically.

Substitute (\ref{k02}) into the Jacobi equation (\ref{f4})
and generate a sequence of
ordinary differential equations
\bea
(H_1w'^{(0)})'-H_2 w^{(0)}=\mu H_3,
\quad   w^{(0)}(t_j)=\eta_j \delta\tau_j^{(0)},
\label{k04}\\
(H_1w'^{(n)})'-H_4 n^2 w^{(n)}-H_2 w^{(n)}=0,
\quad   w^{(n)}(t_j)=\eta_j \delta\tau_j^{(n)}.
\label{k05}
\eea
Thus we recover the inhomogeneous Jacobi equation (\ref{k04})
derived in \cite{FelRub15} for the case of axisymmetric perturbations,
and add a set of homogeneous Jacobi equations (\ref{k05}) for
asymmetric modes.
It is worth to note that solvability conditions for
equations (\ref{k04}, \ref{k05}) with
$\delta\tau_j^{(n)} = 0$ determine the boundary
of the stability region ${\sf C}_n$
for the $n$-th perturbation mode with fixed CL.
The stability analysis described in \cite{FelRub15} for the
axisymmetric perturbations
should be modified and performed for each asymmetric mode independently
to produce the corresponding stability condition (and stability region
${\sf Stab}_n$).
The intersection of all ${\sf Stab}_n$ determines the
stability region ${\sf Stab}$ of the meniscus.

To do this we have to compute the expression for the
second variation $\delta^2W$ given by (\ref{e26}) using
(\ref{k02},\ref{k03}).
First evaluate an expression $\int_{0}^{2\pi}ds [\delta\tau_j(s)]^2$
using Parseval theorem
$$
\int_{0}^{2\pi}\!\!\!\!\!ds
[\delta\tau_j(s)]^2 =
\int_{0}^{2\pi}\!\!\!\!\!ds
\left[\delta\tau_j^{(0)} +
\sum_{n=1}^{\infty}
\delta\tau_j^{(n)} \exp(ins)+c.c.\right]^2 =
\sum_{n=0}^{\infty}|\delta\tau_j^{(n)}|^2
$$
Introducing $\Xi_2^{(0)}[w]$ and $\Xi_2^{(n)}[w]$ for $n > 0$ through
\bea
\Xi_2^{(0)}[w] =
\Xi_2[w^{(0)}] =
2\pi
\int_{t_2}^{t_1}\!\!\!\!\!dt[
H_1(w'^{(0)})^2+H_2(w^{(0)})^2+2\mu H_3 w^{(0)}
],
\nonumber \\
\Xi_2^{(n)}[w] =
\Xi_2[w^{(n)}] =
2\pi
\int_{t_2}^{t_1}\!\!\!\!\!dt[
H_1|w'^{(n)}|^2+n^2H_4|w^{(n)}|^2+H_2|w^{(n)}|^2
],
\nonumber
\eea
we arrive at an expansion
\be
\delta^2W=\sum_{n=0}^{\infty}\delta^2W^{(n)},
\quad
\delta^2W^{(n)} =
\Xi_2[w^{(n)}]+K_1|\delta\tau_1^{(n)}|^2
-K_2|\delta\tau_2^{(n)}|^2.
\label{k07}
\ee

\subsection{Axisymmetric mode stability}
\label{s34}

The complete description of the derivation of the
stability conditions for the axisymmetric mode is given in \cite{FelRub15},
and here we just reproduce the major steps of this approach.

In the general case of free CL one has to find
from (\ref{g7}) the coefficients
$C_1, \; C_2, \; \mu,$ and thus express $\bar{w}^{(0)}(t_j)$
through $\delta\tau_j^{(0)}$.
Multiplying (\ref{k04}) by $\bar{w}^{(0)}(t)$ and integrating
by parts we obtain
$$
\int_{t_2}^{t_1}\left[H_1(\bar{w}^{(0)}_t)^2+H_2(\bar{w}^{(0)})^2\right]dt-
H_1(t)\bar{w}^{(0)}\bar{w}'^{(0)}|_{t_2}^{t_1}=0.
$$
Combining the last equality with (\ref{e26}) we arrive at
\be
\frac{1}{2\pi}\delta^2W^{(0)}=
\frac{1}{2}H_1\bar{w}^{(0)}\bar{w}'^{(0)}|_{t_2}^{t_1}+
K_1[\delta\tau^{(0)}_1]^2-K_2[\delta\tau^{(0)}_2]^2,
\label{g13}
\ee
where $K_j$ are defined in (\ref{e26}). This allows to use only
a part of the solution $\bar{w}^{(0)}(t_j)$ linear in $ \delta\tau^{(0)}_j$
dropping all higher orders.

Write
a general solution $\bar{w}^{(0)}(t)$
of equation (\ref{k04}) built upon the fundamental solutions $\bar{w}^{(0)}_1(t),
\bar{w}^{(0)}_2(t)$ of homogeneous equation, and particular solution of inhomogeneous
equation $\bar{w}^{(0)}_3(t)$,
\be
\bar{w}^{(0)}(t)=C^{(0)}_1\bar{w}^{(0)}_1(t)
+C^{(0)}_2\bar{w}^{(0)}_2(t)+\mu\bar{w}^{(0)}_3(t)\;.
\label{g6}
\ee
Inserting (\ref{g6}) into BC (\ref{e15}) and into constraint (\ref{f2}) we
obtain three linear equations,
\bea
C^{(0)}_1\bar{w}^{(0)}_1(t_j)+C^{(0)}_2\bar{w}^{(0)}_2(t_j)
+\mu\bar{w}^{(0)}_3(t_j)=\bar{w}^{(0)}(t_j),
\
C^{(0)}_1I_1(t_2,t_1)+C^{(0)}_2I_2(t_2,t_1)+\mu I_3(t_2,t_1)=0,
\label{g7}
\eea
where in the expression for $\bar{w}^{(0)}(t_j)=\eta_j \delta\tau^{(0)}_j,$
we retain only the term linear in $\delta\tau^{(0)}_j$ neglecting contributions of
higher orders, and use
$$
I_k(t_2,t_1)=\int_{t_2}^{t_1}dt H_3(t)\bar{w}^{(0)}_k(t).
$$
The case of fixed CL is obtained from (\ref{g7}) by setting
$\bar{w}^{(0)}(t_j) = 0$, and the stability region boundary ${\mathcal C}^{(0)}$
is given by the condition
$\det D^{(0)}(t_2,t_1)= 0,$ where
\be
D^{(0)}(t_2,t_1)=\left(\begin{array}{ccc}
\bar{w}^{(0)}_1(t_2)& \bar{w}^{(0)}_2(t_2)& \bar{w}^{(0)}_3(t_2)\\
\bar{w}^{(0)}_1(t_1)& \bar{w}^{(0)}_2(t_1)& \bar{w}^{(0)}_3(t_1)\\
I_1(t_2,t_1) & I_2(t_2,t_1) & I_3(t_2,t_1)
\end{array}\right).
\label{g8}
\ee
Substituting
the expression for $\bar{w}^{(0)}$
into (\ref{g13}) 
we obtain
\be
\delta^2W^{(0)}=Q^{(0)}_{11}\left[\delta\tau^{(0)}_1\right]^2+
2Q^{(0)}_{12}\delta\tau^{(0)}_1\delta\tau^{(0)}_2+
Q^{(0)}_{22}\left[\delta\tau^{(0)}_2\right]^2.
\label{g14}
\ee

\subsection{Asymmetric mode stability}
\label{s35}
The asymmetric mode stability requires first to find a
solution
$\bar{w}^{(n)}(t)=C^{(n)}_1\bar{w}^{(n)}_1(t)+C^{(n)}_2\bar{w}^{(n)}_2(t),
$
satisfying two boundary conditions
\bea
C^{(n)}_1\bar{w}^{(n)}_1(t_j)+C^{(n)}_2\bar{w}^{(n)}_2(t_j)=
\bar{w}^{(n)}(t_j)=
\eta_j\delta\tau_j^{(n)},
\label{g8a}
\eea
and expressing $\bar{w}^{(n)}(t_j)$
through $\delta\tau_j^{(n)}$.
The case of fixed CL is obtained from (\ref{g8a}) by setting
$\bar{w}^{(n)}(t_j) = 0$. The stability region boundary ${\mathcal C}^{(n)}$
in this case is given by the condition
$\det D^{(n)}(t_2,t_1)= 0,$ where
\be
D^{(n)}(t_2,t_1)=
\left(\begin{array}{cc}
\bar{w}^{(n)}_1(t_2)& \bar{w}^{(n)}_2(t_2)\\
\bar{w}^{(n)}_1(t_1)& \bar{w}^{(n)}_2(t_1)
\end{array}\right).
\label{g8b}
\ee
Multiplying (\ref{k05}) by $\bar{w}^{(n)}(t)$ and integrating
by parts we obtain
$$
\int_{t_2}^{t_1}\left[H_1|\bar{w}^{(n)}_t|^2+
n^2 H_4|\bar{w}^{(n)}|^2+H_2|\bar{w}^{(n)}|^2\right]dt-
H_1(t)|\bar{w}^{(n)}\bar{w}'^{(n)}|_{t_2}^{t_1}=0.
$$
Combining it with (\ref{e26}) we arrive at
\be
\frac{1}{2\pi}\delta^2W^{(n)}=
\frac{1}{2}H_1(t)|\bar{w}^{(n)}\bar{w}'^{(n)}|_{t_2}^{t_1}+
K_1|\delta\tau^{(n)}_1|^2-K_2|\delta\tau^{(n)}_2|^2,
\label{g13a}
\ee
Substituting
the expression for $\bar{w}^{(n)}$
into (\ref{g13}) 
we obtain
\be
\delta^2W^{(n)}=Q^{(n)}_{11}|\delta\tau^{(n)}_1|^2+
2Q^{(n)}_{12}|\delta\tau^{(n)}_1||\delta\tau^{(n)}_2|+
Q^{(n)}_{22}|\delta\tau^{(n)}_2|^2.
\label{g14a}
\ee
The
necessary conditions to have $\delta^2W^{(n)}\geq 0$ are given by three inequalities,
\bea
Q^{(n)}_{11}(t_2,t_1)\geq 0,
\quad Q^{(n)}_{22}(t_2,t_1)\geq 0,
\quad Q^{(n)}_{33}(t_2,t_1)=
Q^{(n)}_{11}Q^{(n)}_{22}-[Q^{(n)}_{12}]^2 \geq 0,
\label{g15a}
\eea
Recalling the expression (\ref{k07}) for the second variation
$\delta^2W$ we see that
\be
\delta^2W=\sum_{n=0}^{\infty}\delta^2W^{(n)}
=\sum_{n=0}^{\infty}
\left[Q^{(n)}_{11}|\delta\tau^{(n)}_1|^2+
2Q^{(n)}_{12}|\delta\tau^{(n)}_1||\delta\tau^{(n)}_2|+
Q^{(n)}_{22}|\delta\tau^{(n)}_2|^2
\right].
\label{g30}
\ee
Due to arbitrariness of $\delta\tau_j$, it follows from (\ref{g30})
one has to require the stability of the each mode independently of the
others, so that the condition $\delta^2W^{(n)}\geq 0$ should hold
for every $n$.
The boundary ${\mathcal B}^{(n)}$ of the stability
region ${\sf Stab}^{(n)}$ of the $n$-th mode is given by the
simultaneous equalities in (\ref{g15a}). It should be underlined that
the ${\sf Stab}^{(n)}$ should lie inside the region ${\sf C}$
bounded by the intersection of all ${\sf C}_n$.



\section{Computation of $Q^{(n)}_{ii}$}
\label{s4}

The computation of the explicit expressions for $Q_{ii}$ can be split into two
independent steps -- first, evaluate $K_j$, and, second,
find the solutions $\bar w^{(n)},$ and their derivatives $\bar w'^{(n)}$.

\subsection{Computation of $K_j$}
\label{s41}

Find the explicit expression for $K_j$ in (\ref{e31}).
The matrix ${\bf F_{tr}}$ can be presented as
$
{\bf F_{tr}} =
|{\bf t}|^{-1}\;{\bf e}_r\otimes{\bf t}-
S_H r\;{\bf e}_r\otimes{\bf e}_z,
$
where ${\bf e}_r$ and ${\bf e}_z$ denote the unit vectors in the
$r$ and $z$ direction, respectively.
First find
$$
\frac{\partial F_j}{\partial {\bf t}} +
\frac{\partial G_j}{\partial {\bf T}_j} =
R_j\left(
\bar{{\bf t}}_j-
\frac{\langle\bar{{\bf t}}_j,{\bf T}_j\rangle{\bf T}_j}
{\langle{\bf T}_j,{\bf T}_j\rangle}
\right),
\
\frac{\partial G_j}{\partial {\bf R}_j} =
\left(S_H R_j Z_j'-
\langle\bar{{\bf t}}_j,{\bf T}_j\rangle
\right){\bf e}_r.
$$
Find the term related to ${\bf F_{tr}}$ in the expression (\ref{e31}), it reads
$$
\langle {\bf T}_j,{\bf F_{tr}}(t_j)\cdot{\bf T}_j\rangle=
-R_j'\left(S_H R_j Z_j'-
\langle\bar{{\bf t}}_j,{\bf T}_j\rangle
\right),
$$
and we obtain
$$
K_j =
\frac{R_j}{2|\bar{{\bf t}}_j|}\left(
\langle\bar{{\bf t}}_j,{\bf T}'_j\rangle
-\frac{\langle\bar{{\bf t}}_j,{\bf T}_j\rangle
\langle{\bf T}_j,{\bf T}'_j\rangle}
{\langle{\bf T}_j,{\bf T}_j\rangle}
\right)-
\frac{H_1(t_j)}{2}\eta_j\eta_j'.
$$
Using the definitions of the
normal to the SB: ${\bf N}_1=\{Z_1',-R_1'\}, \;
{\bf N}_2=-\{Z_2',-R_2'\},$
the expression in the round brackets can be written as
$$
\langle\bar{{\bf t}}_j,{\bf T}'_j\rangle
-\frac{\langle\bar{{\bf t}}_j,{\bf T}_j\rangle
\langle{\bf T}_j,{\bf T}'_j\rangle}
{\langle{\bf T}_j,{\bf T}_j\rangle} =
(-1)^{j}\frac{\langle{\bf N}_j,{\bf T}'_j\rangle}
{\langle{\bf T}_j,{\bf T}_j\rangle}\langle{\bf n}_j,{\bf T}_j\rangle =
(-1)^{j+1}\frac{\langle{\bf N}'_j,{\bf T}_j\rangle}
{\langle{\bf T}_j,{\bf T}_j\rangle}\;\eta_j.
$$
Collecting all terms %
we arrive at
\be
K_j = -\frac{\eta_jR_j}{2}V_j,\
V_j =
(-1)^{j}
\frac{\langle{\bf N}'_j,{\bf T}_j\rangle}
{\langle{\bf T}_j,{\bf T}_j\rangle}+
\eta_j' =
\left\langle
\frac{(-1)^{j}{\bf N}'_j}
{\langle{\bf T}_j,{\bf T}_j\rangle}+
{\bf n}'_j,
{\bf T}_j
\right\rangle.
\label{K_j}
\ee
Using the definition of
the vectors ${\bf N}_j,{\bf T}_j,$ we obtain
\be
V_j = R_j'z_j''-Z_j'r_j''-\frac{R_j'Z_j''-Z_j'R_j''}{R_j'^2+Z_j'^2}=
\eta'_j - \tilde{V}_j.
\label{V_j}
\ee

\subsection{Computation of $\bar w^{(n)}$}
\label{s42}

The inhomogeneous Jacobi equation (\ref{k04}) reads
\be
(rw'^{(0)})'r'-\left(rr''\right)'w^{(0)} =\mu rr'.
\label{h5b}
\ee
Here $r(t)=\sqrt{1+B^2+2B\cos S_Ht},$ denotes
a solution of the YLE describing both unduloids ($B<1$), and
nodoids ($B>1$), as well as cylinder ($B=0$) and sphere ($B=1$).
The nodoids may exist of two types -- convex with $S_H=1$ and
concave with $S_H=-1$. The solution for $z(t)$ is expressed through
the elliptic integrals of the first and second kind (see \cite{FelRub15, Slob1982})
and satisfies a relation $r'^2+z'^2=1$.

It is easy to check by the direct computation that the
homogeneous Jacobi equation with $\mu=0$ has a solution $\bar w^{(0)}_1=r'$,
while the second solution reads $\bar w^{(0)}_2=\bar w^{(0)}_1U$,
where $rr'^2U'=1$.
It can be shown that $\bar w^{(0)}_2$ as well
the solution of the inhomogeneous problem $\bar w^{(0)}_3$
can be expressed through the elliptic integrals
of the first and second kind (see \cite{FelRub15, Slob1982})
\bea
&&\bar w^{(0)}_1=r',
\quad
\bar w^{(0)}_2=\cos t+(1+B)M_1\bar{w}^{(0)}_1,
\quad
\bar w^{(0)}_3=1+(1+B)M_2\bar{w}^{(0)}_1,
\label{h5d} \\
&&
M_1(t,m)=E(t/2,m)-F(t/2,m)+M_2,
\
M_2(\phi,m)=m^2F(t/2,m)/2,
\ m=2\sqrt{B}/(1+B).
\nonumber
\eea
The homogeneous Jacobi equation (\ref{k05}) reads
\be
(rw'^{(n)})'rr'-\left(rr''\right)'rw^{(n)}-n^2r'w^{(n)}=0.
\label{h5c}
\ee
It is easy to check by direct computation that for $n=1$
this equation has a solution $\bar w^{(1)}_2=z'$ (see \cite{Myshkis87, Slob1982}),
where $r'^2+z'^2=1$,
and $\bar w^{(1)}_1$ again is
expressed through the elliptic integrals
\be
\bar w^{(1)}_1=-B \sin t+[(1+B)E(t/2,m)+(1-B)F(t/2,m)]\bar{w}^{(1)}_2=rr'+zz',
\quad
\bar w^{(1)}_2=z'.
\label{h5e}
\ee
The general analytical solutions
$\bar w^{(n)}_k$ for $n>1$ are not known.
In the particular case $B=n$ one has $\tilde w^{(n)}_j = \bar w^{(n)}_j|_{B=n}$
and finds:
\be
\tilde w^{(n)}_1=-\sin t+(1+n)[(1+n)E(t/2,m)-(1-n)F(t/2,m)]\tilde{w}^{(n)}_2,
\quad
\tilde w^{(n)}_2=\frac{n+\cos t}{r}.
\label{h5e1}
\ee
In all three cases the solutions satisfy the
following conditions
$w_1(0)=0,\; w'_1(0)=const > 0,$ and
$w_2(0)=const > 0,\; w'_2(0)=0$.
In Appendix \ref{appendix0} we perform the analysis of the
Jacobi equation (\ref{k05}) and show how to obtain the
fundamental solutions described above.

\subsection{Computation of $\bar w'^{(n)}$}
\label{s43}
The computation of the first derivative $\bar w'^{(n)}(t_j)$ at the
end points $t_j$ is straightforward and we present here the main
steps and the final result.
The case of axisymmetric mode should be considered separately, and
we examine it first.

Use the conditions (\ref{g7}) to find the constants
$C^{(0)}_1,C^{(0)}_2$ and $\mu$.
Introduce two determinants $B_j(t)$ 
\be
A^{(0)}_1(t)=\left|\begin{array}{ccc}
\bar{w}^{(0)}_1(t_2)&\bar{w}^{(0)}_2(t_2)&\bar{w}^{(0)}_3(t_2)\\
\bar{w}^{(0)}_1(t) & \bar{w}^{(0)}_2(t)& \bar{w}^{(0)}_3(t)\\
I_1 & I_2 & I_3\end{array}\right|,\quad
A^{(0)}_2(t)=\left|\begin{array}{ccc}
\bar{w}^{(0)}_1(t)&\bar{w}^{(0)}_2(t)&\bar{w}^{(0)}_3(t)\\
\bar{w}^{(0)}_1(t_1) & \bar{w}^{(0)}_2(t_1)& \bar{w}^{(0)}_3(t_1)\\
I_1 & I_2 & I_3\end{array}\right|,
\label{s43e1}
\ee
Direct computation shows that
\bea
\bar{w}^{(0)}(t_1)\bar{w}'^{(0)}(t_1) &=&
\frac{\eta_1^2A'^{(0)}_1(t_1)[\delta\tau^{(0)}_1]^2
+\eta_1\eta_2A'^{(0)}_2(t_1)\delta\tau^{(0)}_1\delta\tau^{(0)}_2}
{A^{(0)}_1(t_1)},
\nonumber \\
\bar{w}^{(0)}(t_2)\bar{w}'^{(0)}(t_2) &=&
\frac{\eta_2^2A'^{(0)}_2(t_2)[\delta\tau^{(0)}_2]^2
+\eta_1\eta_2A'^{(0)}_1(t_2)\delta\tau^{(0)}_1\delta\tau^{(0)}_2}
{A^{(0)}_2(t_2)}.
\label{s43e2}
\eea

The case of arbitrary asymmetric mode is considered similarly.
First, we use the boundary conditions (\ref{g8a}) and
find the expressions for $C^{(n)}_1$ and $C^{(n)}_2$.
Then we introduce two determinants
$A^{(n)}_j(t)$ through the relations
$$
A^{(n)}_1(t)=\left|\begin{array}{cc}
\bar{w}^{(n)}_1(t_2)&\bar{w}^{(n)}_2(t_2)\\
\bar{w}^{(n)}_1(t) & \bar{w}^{(n)}_2(t)
\end{array}\right|,
\quad
A^{(n)}_2(t)=\left|\begin{array}{cc}
\bar{w}^{(n)}_1(t)&\bar{w}^{(n)}_2(t)\\
\bar{w}^{(n)}_1(t_1) & \bar{w}^{(n)}_2(t_1)
\end{array}\right|,
$$
Simple algebra shows that
\bea
\bar{w}^{(n)}(t_1)\bar{w}'^{(n)}(t_1) &=&
\frac{\eta_1^2A'^{(n)}_1(t_1)|\delta\tau^{(n)}_1|^2
+\eta_1\eta_2A'^{(n)}_2(t_1)|\delta\tau^{(n)}_1||\delta\tau^{(n)}_2|}
{A^{(n)}_1(t_1)},
\nonumber \\
\bar{w}^{(n)}(t_2)\bar{w}'^{(n)}(t_2) &=&
\frac{\eta_2^2A'^{(n)}_2(t_2)|\delta\tau^{(n)}_2|^2
+\eta_1\eta_2A'^{(n)}_1(t_2)|\delta\tau^{(n)}_1||\delta\tau^{(n)}_2|}{A^{(n)}_2(t_2)},
\label{s43e4}
\eea
where $A^{(n)}_1(t_1)=A^{(n)}_2(t_2)=A^{(n)}$.
It is clear that (\ref{s43e4}) includes (\ref{s43e2}) as
a particular case for $n=0$.

\subsection{Computation of $Q^{(n)}_{ij}$}
\label{s44}
Substitution of (\ref{s43e4}) into
(\ref{g14}, \ref{g30}) produces
\bea
Q^{(n)}_{jj} &=&
(-1)^{j+1}
\left[
K_j +
\frac{\eta_j^2H_1(t_j)}{2}\frac{A'^{(n)}_j(t_j)}{A^{(n)}}
\right],
\quad
j = 1,2,
\label{s44e01} \\
Q^{(n)}_{12} &=&
\frac{\eta_1\eta_2}{2}
\frac{H_1(t_1)A'^{(n)}_2(t_1)
}{A^{(n)}}
=-
\frac{\eta_1\eta_2}{2}
\frac{H_1(t_2)A'^{(n)}_1(t_2)
}{A^{(n)}}.
\label{s44e02}
\eea
Using the expression (\ref{K_j}) for $K_j$ we
write explicit representation of $Q^{(n)}_{ij}$
\bea
Q^{(n)}_{jj} &=&
(-1)^{j+1}\frac{\eta_j^2R_j}{2A^{(n)}}
\left[
-(V_j/\eta_j) A^{(n)}+ A'^{(n)}_j(t_j)
\right],
\label{s44e1} \\
Q^{(n)}_{12} &=&
\frac{\eta_1\eta_2 R_1A'^{(n)}_2(t_1)}{2A^{(n)}}
=-\frac{\eta_1\eta_2 R_2A'^{(n)}_1(t_2)}{2A^{(n)}}.
\label{s44e2}
\eea
The condition $Q_{jj}=0$ is satisfied either by setting $\eta_j=0$
(which corresponds to the meniscus existence boundary,
see \cite{RubFel2015}), or by requiring
$A'^{(n)}_j(t_j) - (V_j/\eta_j)A^{(n)}_j(t_j)=0$.
The last relation is equivalent to an inhomogeneous linear BC
on the $n$-th mode perturbation at the end points of the interval
\be
(V_j/\eta_j)w^{(n)}(t_j)-w'^{(n)}(t_j)=0.
\label{s44e3}
\ee
As this BC is valid for
every perturbation mode it implies that the same condition
should be met for an arbitrary asymmetric perturbation
(valid for nonzero $\eta_j$, i.e., everywhere in the existence region):
\be
(V_j/\eta_j)w(t_j)-w'(t_j)=0.
\label{s44e4}
\ee
In Appendix \ref{appendix1} we show that
$V_j/\eta_j = (-1)^{j+1}\chi_j,$ where the quantity $\chi_j$ was
introduced in \cite{Myshkis87}, Ch.3. Then the conditions (\ref{s44e4})
reduce to
$$
\chi_1 w(t_1)-w'(t_1)=0,
\quad
\chi_2 w(t_2)+w'(t_2)=0.
$$
The expression for
$Q^{(n)}_{33}=Q^{(n)}_{11}Q^{(n)}_{22}-Q^{(n)}_{12}Q^{(n)}_{21},$ reads
$$
Q^{(n)}_{33} =
R_1R_2\left[\frac{\eta_1\eta_2}{2A^{(n)}}
\right]^2
\left\{
A'^{(n)}_2(t_1)A'^{(n)}_1(t_2)-
[A'^{(n)}_1(t_1)-(V_1/\eta_1)A^{(n)}]
[A'^{(n)}_2(t_2)-(V_2/\eta_2)A^{(n)}]
\right\}.
$$
Thus, the condition $Q^{(n)}_{33}=0,$ which determines the
stability region boundary ${\mathcal B}^{(n)}$ is written as
\be
\left[-(V_1/\eta_1)A^{(n)}+A'^{(n)}_1(t_1)\right]
\left[-(V_2/\eta_2)A^{(n)}+A'^{(n)}_2(t_2)\right]-
A'^{(n)}_2(t_1)A'^{(n)}_1(t_2)=0.
\label{s44e5}
\ee
Introduce two determinants
$$
A^{(0)}_3=\left|\begin{array}{ccc}
\bar{w}'^{(0)}_1(t_2)&\bar{w}'^{(0)}_2(t_2)&\bar{w}'^{(0)}_3(t_2)\\
\bar{w}'^{(0)}_1(t_1) & \bar{w}'^{(0)}_2(t_1)& \bar{w}'^{(0)}_3(t_1)\\
I_1 & I_2 & I_3
\end{array}\right|,
\quad
A^{(n)}_3=
\left|\begin{array}{cc}
\bar{w}'^{(n)}_1(t_2)&\bar{w}'^{(n)}_2(t_2)\\
\bar{w}'^{(n)}_1(t_1) & \bar{w}'^{(n)}_2(t_1)\\
\end{array}\right|.
$$
Direct computation shows that the following relation holds:
$$A^{(n)}_3 A^{(n)} =
A'^{(n)}_1(t_1)A'^{(n)}_2(t_2)-A'^{(n)}_1(t_2)A'^{(n)}_2(t_1).$$
Using it we rewrite (\ref{s44e5})
\be
V_1V_2A^{(n)}-
V_1\eta_2A'^{(n)}_2(t_2)-
V_2\eta_1A'^{(n)}_1(t_1)+
\eta_1\eta_2A^{(n)}_3=0.
\label{s44e6}
\ee

\subsection{Relations between conditions $Q^{(n)}_{ii}=0$}
\label{s45}

Consider the BC (\ref{s44e3}) and use the representation
of the perturbation modes (\ref{g6}) for $n=0$ and (\ref{g8a})
for $n>0$, respectively. For the axisymmetric mode we
find the solvability condition for (\ref{s44e3}) as
vanishing determinant
\be
\!\!\!D_M^{(0)}\!=\!\left(\begin{array}{ccc}
V_2\bar{w}^{(0)}_1(t_2)/\eta_2-\bar{w}'^{(0)}_1(t_2)&
V_2\bar{w}^{(0)}_2(t_2)/\eta_2-\bar{w}'^{(0)}_2(t_2)&
V_2\bar{w}^{(0)}_3(t_2)/\eta_2-\bar{w}'^{(0)}_3(t_2)\\
V_1\bar{w}^{(0)}_1(t_1)/\eta_1-\bar{w}'^{(0)}_1(t_1)&
V_1\bar{w}^{(0)}_2(t_1)/\eta_1-\bar{w}'^{(0)}_2(t_1)&
V_1\bar{w}^{(0)}_3(t_1)/\eta_1-\bar{w}'^{(0)}_3(t_1)\\
I_1 & I_2 & I_3
\end{array}\right).
\label{p01}
\ee
Direct computation shows that the condition $\det D_M^{(0)}=0$
coincides with (\ref{s44e6}) for $n=0$.
Similarly, introducing a condition
\be
\!\!\!\det D_M^{(n)}\!=\!\left|\begin{array}{cc}
V_2\bar{w}^{(n)}_1(t_2)/\eta_2-\bar{w}'^{(n)}_1(t_2)&
V_2\bar{w}^{(n)}_2(t_2)/\eta_2-\bar{w}'^{(n)}_2(t_2)\\
V_1\bar{w}^{(n)}_1(t_1)/\eta_1-\bar{w}'^{(n)}_1(t_1)&
V_1\bar{w}^{(n)}_2(t_1)/\eta_1-\bar{w}'^{(n)}_2(t_1)
\end{array}\right|=0,
\label{p02}
\ee
we find that it
coincides with (\ref{s44e6}) for $n>0$.

This observation implies that the BC (\ref{e15}) with arbitrary
$\delta\tau_j$ are consistent with the conditions (\ref{s44e4}).
It also means that the stability boundary ${\mathcal B}^{(n)}$
for the $n$-th perturbation mode is determined
solely by the condition $Q_{33}^{(n)}=0$.

\section{Computation of stability regions}
\label{s6}

From the computational point of view, the
determination of the stability region ${\sf Stab}$ requires
first
to determine all regions of stability ${\sf C}_n$ for the fixed CL bounded by
${\mathcal C}^{(n)}$ and find their intersection
${\sf C}=\cap_{n=0}^{\infty} {\sf C}_n$.
Then
for each $n \ge 0$ find ${\sf Stab}_n$ bounded by
${\mathcal B}^{(n)}$ which lies within ${\sf C}$, and
obtain ${\sf Stab}=\cap_{n=0}^{\infty} {\sf Stab}_n$.

\subsection{Stability region boundary for menisci with fixed CL}
\label{s61}

The boundary ${\mathcal C}^{(0)}$ is specified by the condition
$\det D^{(0)} = 0,$ where the matrix $D^{(0)}$ is given
in (\ref{g8}), and its elements presented in (\ref{h5d}).
For $n > 0$, the relation
$\det D^{(n)} = 0$  defines the boundary ${\mathcal C}^{(n)}$
where the matrix $D^{(n)}$ is given in (\ref{g8b}).
It can be written as
\be
A^{(n)}=\bar{w}^{(n)}_1(t_1)\bar{w}^{(n)}_2(t_2)-
\bar{w}^{(n)}_1(t_2)\bar{w}^{(n)}_2(t_1)=0,
\label{eq7}
\ee
which implicitly defines a curve in the plane $\{t_1,t_2\}$.
In Appendix \ref{appendix2} we discuss a computational procedure
establishing the curve ${\mathcal C}^{(n)}$ and show
that the boundary ${\mathcal C}^{(1)}$ exists only
for nodoids ($B > 1$).

\begin{figure}[h!]
\begin{center}
\begin{tabular}{cc}
rm F	\psfig{figure=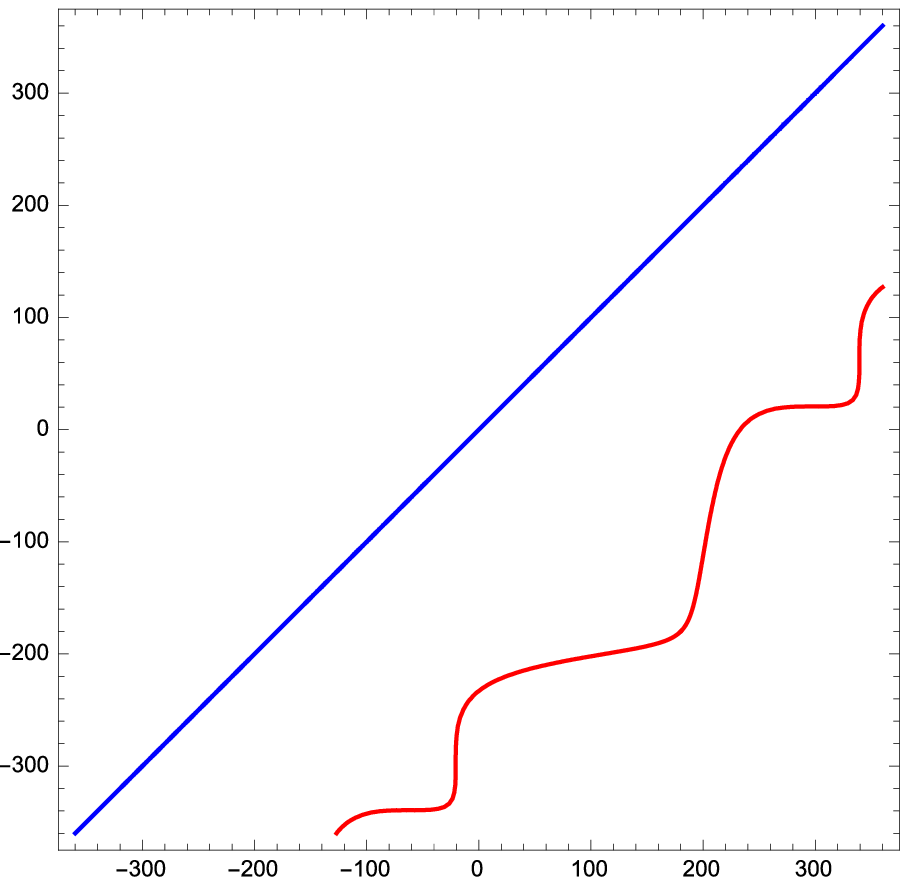,width=5.5cm}&
\psfig{figure=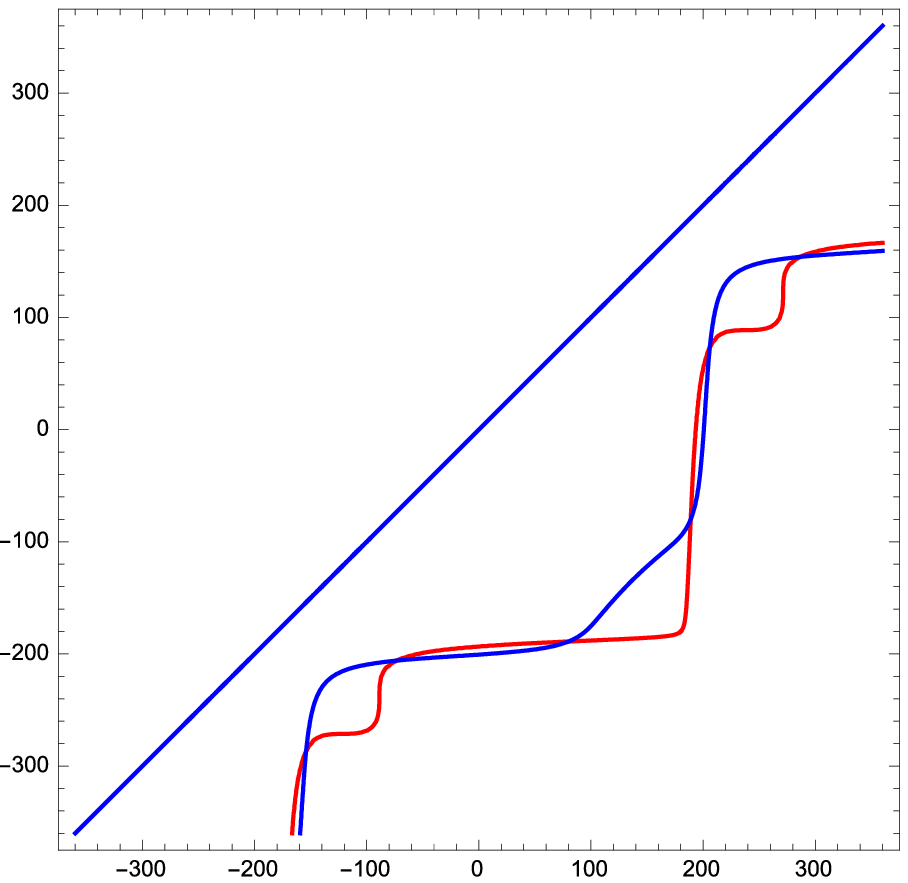,width=5.5cm}\\
(a) & (b)\\
\psfig{figure=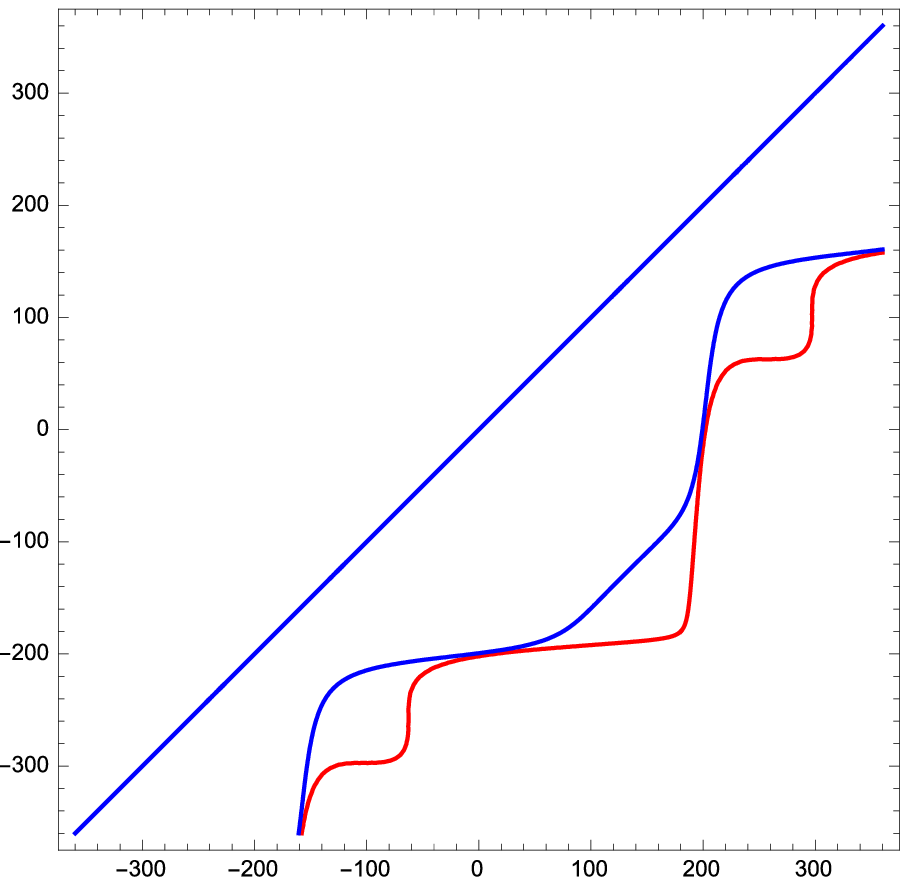,width=5.5cm}&
\psfig{figure=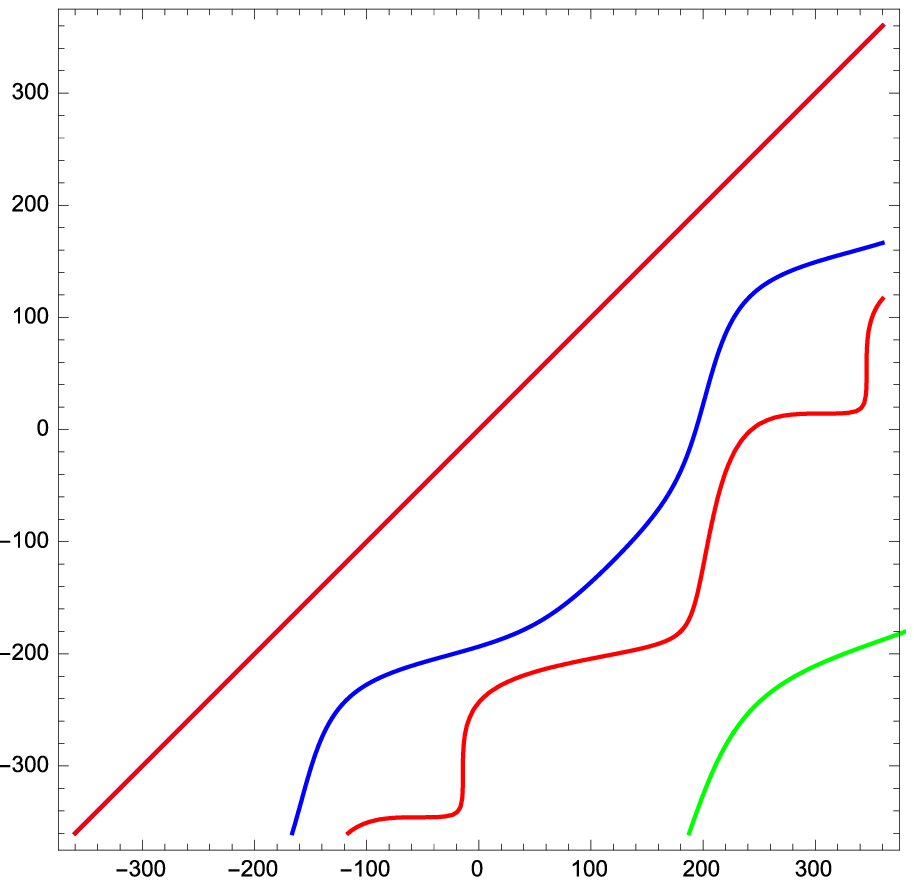,width=5.5cm}\\
(c) & (d)\\
\psfig{figure=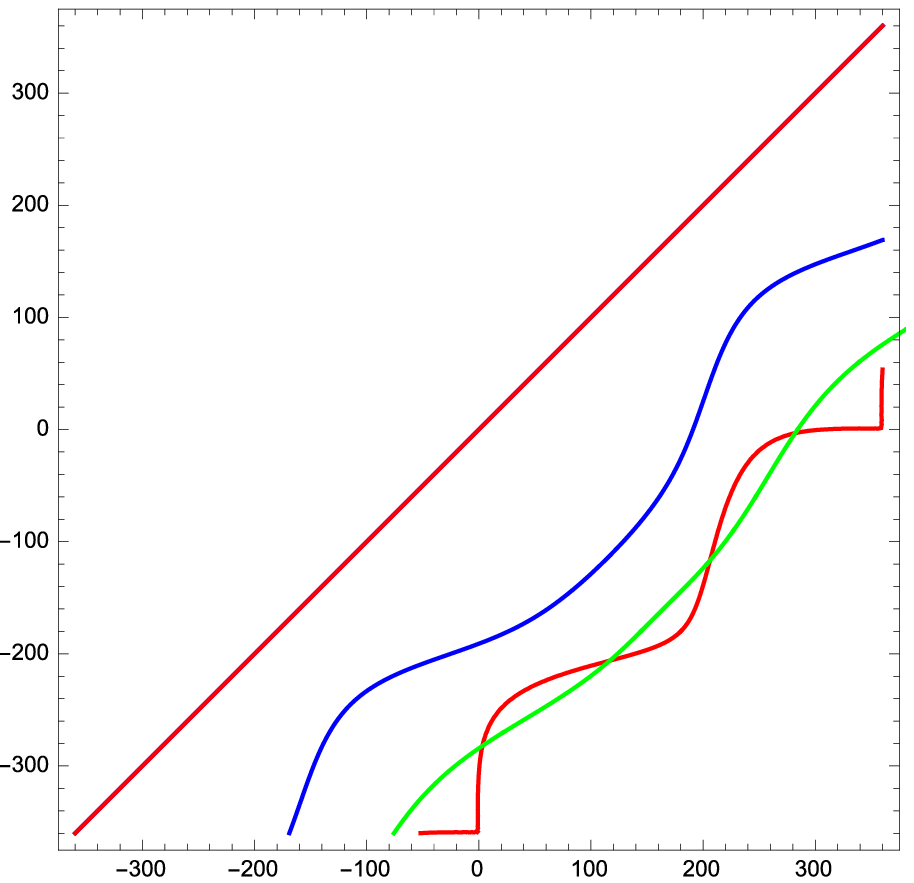,width=5.5cm}&
\psfig{figure=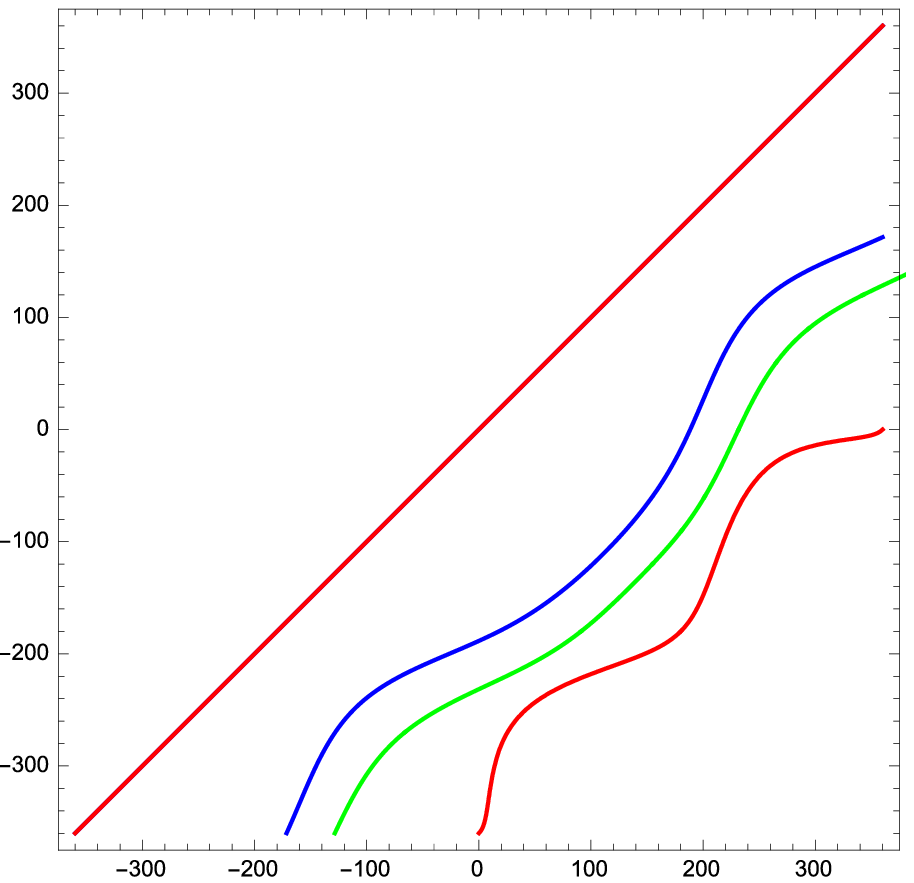,width=5.5cm}\\
(e) & (f)
\end{tabular}
\end{center}
\caption{The boundaries ${\mathcal C}^{(n)}$ of the stability regions ${\sf C}_n$
for fixed CL for
$n=0$ (red), $n=1$ (blue), $n=2$ (green), and
a) $B=0.5$, b) $B=1.4$, c) $B=\pi/2$, d) $B=2.1$, e) $B=2.4$, f) $B=2.8$.
}
\label{f02}
\end{figure}

For $n > 1$ the boundary ${\mathcal C}^{(n)}$ must be computed numerically.
Numerical simulations show that the boundary ${\mathcal C}^{(n)}$ of the
$n$-th perturbation mode exists for $B > n$ only.
This means that for unduloids ($0<B<1$) the only restriction imposed by
the fixed CL is given by ${\mathcal C}^{(0)}$, while for the
nodoids with $B>1$ the boundaries ${\mathcal C}^{(n)}$ with $n>0$ may
reduce the stability region. First, we checked relative position of the
boundaries ${\mathcal C}^{(0)}$ and ${\mathcal C}^{(1)}$ for $1 < B < 2$.
We found that for $1 < B < \pi/2$ these curves intersect, while for
$B > \pi/2$ the curve ${\mathcal C}^{(1)}$ lies inside the region ${\sf C}_0$
(see Figure \ref{f02}).
For $B > 2$ we checked the influence of ${\mathcal C}^{(2)}$
on the shape of the stability region, and find out that it always lies
outside of ${\sf C}_1$. The relative position of
between ${\mathcal C}^{(0)}$ and ${\mathcal C}^{(2)}$ changes with $B$, namely,
for $B$ values close to $2$ we observe ${\mathcal C}^{(2)}$ outside of
${\sf C}_0$, but with growth of $B$ is approaches ${\mathcal C}^{(0)}$,
then intersects it and then ${\mathcal C}^{(2)}$ is completely
between ${\mathcal C}^{(0)}$ and ${\mathcal C}^{(1)}$.

Thus, the numerical analysis implies that the stability region ${\sf C}$
for nodoids with fixed CL for $1 < B < \pi/2$ is determined by interplay of the
boundaries ${\mathcal C}^{(0)}$ and ${\mathcal C}^{(1)}$, while
for larger values of $B$ it is completely defined by ${\mathcal C}^{(1)}$ only.

\subsection{Stability region boundary for menisci with free CL}
\label{s62}

Turning to computation of the stability region for the
menisci with free CL between two axisymmetric solid bodies
one has first to establish the region of existence for the given meniscus
(i.e., given values of $B$ and $S_H$) and
the given SB (i.e., given ${\bf R}_j$). This region
${\sf Exist}(B,S_H,{\bf R}_1,{\bf R}_2)$
is determined by a set of conditions
(some of them are discussed in details in \cite{RubFel2015}).
Then the construction of the boundaries ${\mathcal B}^{(n)}$
should be done only inside the existence region.

The method developed in \cite{Myshkis87} states that in order
to establish the meniscus stability w.r.t. asymmetric perturbations
it is sufficient to determine the boundary ${\mathcal B}^{(1)}$ of the
first mode ($n=1$) only, except the case of the meniscus between two
parallel plates when the boundary ${\mathcal B}^{(2)}$ for $n=2$ also should be
taken into account.
We start with this particular case.

\subsubsection{Two parallel plates}
\label{s621}
It is easy to check that in this case $Z'=Z''=R''=0,$ and $R'=1,$
so that we find $\eta_j = z'_j,$ and $V_j = \eta'_j$. The condition
(\ref{s44e3}) reduces to
$$
\eta_j w'^{(n)}(t_j)= \eta'_j w^{(n)}(t_j),
\quad
w^{(n)}(t_j) = \eta_j = z'_j.
$$
Substitute it into (\ref{h5c}) we obtain
$$
(r \eta')'-\frac{n^2}{r}\eta-\frac{(rr'')'}{r'}\eta=0.
$$
Note that $\eta = z'$ identically satisfies equation (\ref{h5c}) with $n=1$.
This means that the first mode boundary ${\mathcal B}^{(1)}$ does not exist, while
${\mathcal B}^{(n)}$ for $n>1$ should satisfy
an meniscus existence condition
$\eta_j = z'(t_j) = 0,$ mentioned above. Using the explicit expression for
$z'(t)=(1+B \cos t)/r,$ we find the
boundaries $t_j=t^*,$ where $\cos t^* = -1/B$.

This result shows that the stability regions
${\sf Stab}_0$ found in \cite{FelRub15} for unduloids between two
parallel plates coincide with the stability regions
${\sf Stab}$ valid for arbitrary asymmetric perturbations.
It also indicates that the boundary ${\mathcal B}^{(n)}$ of the
stability region for the asymmetric perturbations
exist only for nodoids ($B > 1$), and this boundary
coincides with the existence boundary of nodoids between two
parallel plates. Thus, in this case the stability region ${\sf Stab}$
is determined by intersection of the stability region
of the axisymmetric perturbation and the stability regions for
asymmetric perturbation modes with fixed CL: ${\sf Stab} =
{\sf Stab}_0 \cap {\sf C}$.

\begin{figure}[h!]
\begin{center}
\begin{tabular}{cc}
\psfig{figure=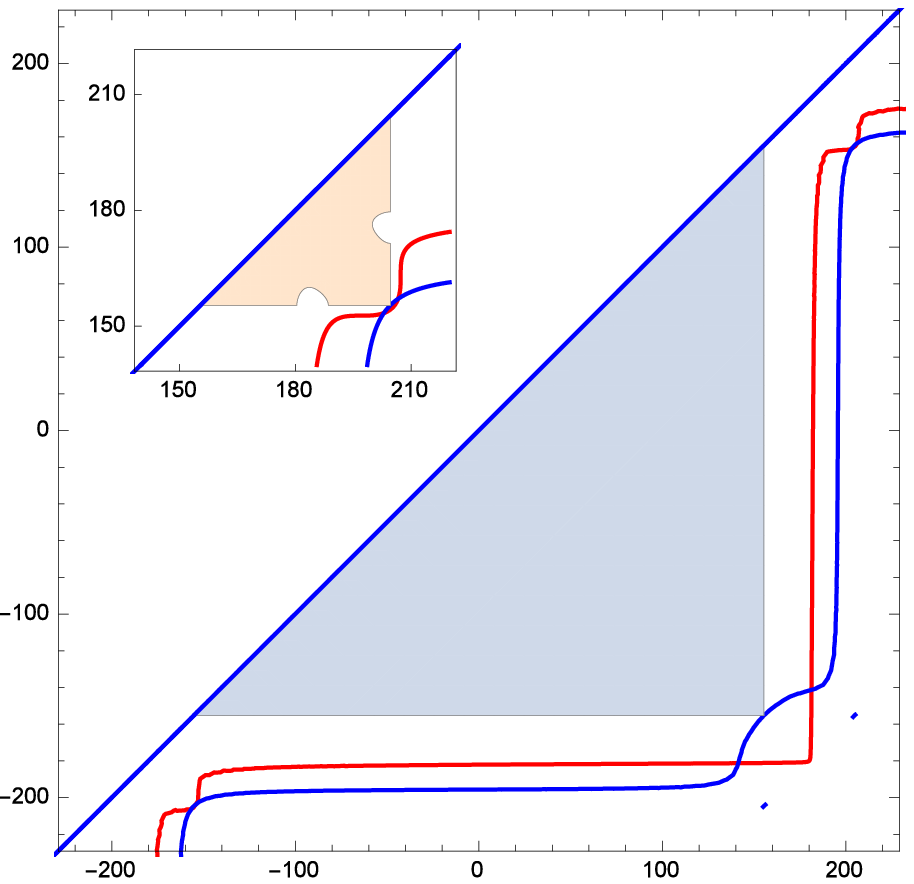,width=6.0cm}&
\psfig{figure=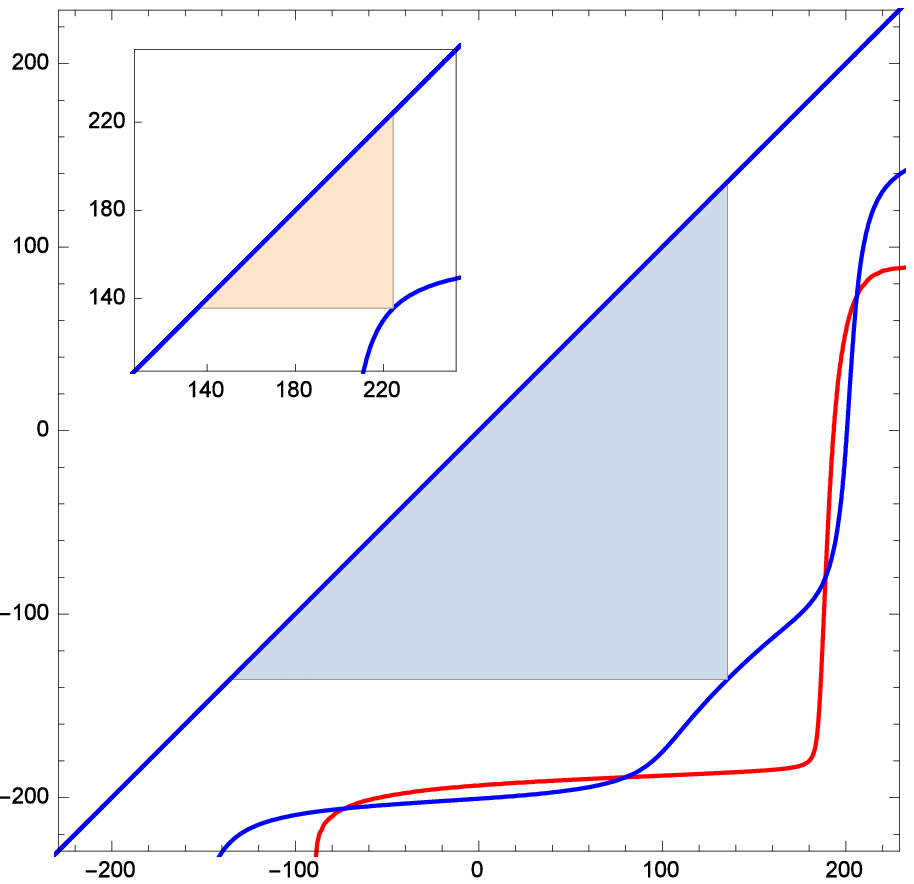,width=6.0cm}\\
(a) & (b)
\end{tabular}
\end{center}
\caption{The stability region for nodoids with a) $B=1.1$
and b) $B=1.4$ between parallel plates.
The shaded areas determine the stability region w.r.t.
axisymmetric perturbations for the convex ($S_H=1$, blue) and
concave ($S_H=-1$, orange) nodoids.
Solid curves represent fixed CL stability boundary
${\mathcal C}^{(n)}$ for $n=0$ (red)
and $n=1$ (blue).}
\label{f06}
\end{figure}
The computations nevertheless show that ${\sf Stab} =
{\sf Stab}_0$ (see Figure \ref{f06});
the boundary ${\mathcal C}^{(1)}$ only {\it touches}
the region ${\sf Stab}_0$, but never intersects it.
The contact point of ${\mathcal C}^{(1)}$ and ${\sf Stab}_0$
for the convex [concave] nodoid shown in Figure \ref{f06}
is given by $t_2=t^*,\;t_1=t^*[2\pi-t^*],$ when the matrix $D^{(1)}(t_2,t_1)$
is degenerate.

\subsection{Influence of asymmetric perturbations on stability region}
\label{s63}
In \cite{RubFel2015} the stability regions for the
axisymmetric menisci under axisymmetric perturbations
were established for various geometrical settings.
It is instructive to figure out how asymmetric perturbations
affect these stability regions.

The condition (\ref{s44e3}) leads to the explicit expression for
the stability boundary ${\mathcal B}^{(1)}$ of the first asymmetric perturbation mode
\be
C_{11}C_{22}-C_{12}C_{21} = 0,
\quad
C_{ij} =
\eta_j w'^{(1)}_i(t_j) - V_j w^{(1)}_i(t_j),
\label{eq9}
\ee
where $w^{(1)}_i$ are given by (\ref{h5e}).
In Appendix \ref{appendix3} we discuss a computational procedure determining
the boundary ${\mathcal B}^{(n)}$ for asymmetric modes with $n > 1$.
The approach used in \cite{Myshkis87} implies that
in order to find the stability region ${\sf Stab}_1$ for asymmetric perturbations
it is enough to consider only a part of the boundary ${\mathcal B}^{(1)}$
that lies inside ${\sf C}_1$.
Numerical simulations show that
the boundary ${\mathcal B}^{(1)}$
in some cases might exist for arbitrary positive $B$.
This means that both
unduloid and nodoid stability regions might be reduced by
asymmetric perturbations. Nevertheless, we did not
find any combinations of the parameters for which
the boundary ${\mathcal B}^{(1)}$ crosses the
stability region for axisymmetric perturbations.
The same time the boundary ${\mathcal C}^{(1)}$ does
reduce the stability region of nodoid menisci with $B > 1$.
As an example we discuss below the
stability of the nodoid menisci between two solid spheres.

\subsubsection{Two equal spheres}
\label{s622}
For two spheres of the same radius $a$ we have
$R_j = a \sin\tau_j,\; Z_j = (-1)^{j} a \cos\tau_j,$ where
the angles $\tau_j$ parameterize the
spherical surfaces and are found from the condition $R_j = r_j,$
i.e., $a \sin\tau_j = \sqrt{1+B^2+2B\cos S_H t_j}$.
It is easy to obtain the following relations:
$$
\eta_j =
\frac{aS_H}{r_j}
\left[
\cos\tau_j + B\cos(S_H t_j +(-1)^{j+1} \tau_j)
\right],
\quad
\tilde{V}_j = (-1)^{j+1}.
$$
Substitution of these expressions
and the solutions (\ref{h5e}) into (\ref{eq9}) produces
an explicit condition for the boundary ${\mathcal B}^{(1)}$.
We found that in some cases
${\mathcal B}^{(1)}$ can intersect ${\mathcal C}^{(1)}$,
but it happens outside of the existence region.
On the contrary, the curve ${\mathcal B}^{(1)}$ never crossed
${\sf Stab}_0$.

\begin{figure}[h!]
\begin{center}
\begin{tabular}{cc}
\psfig{figure=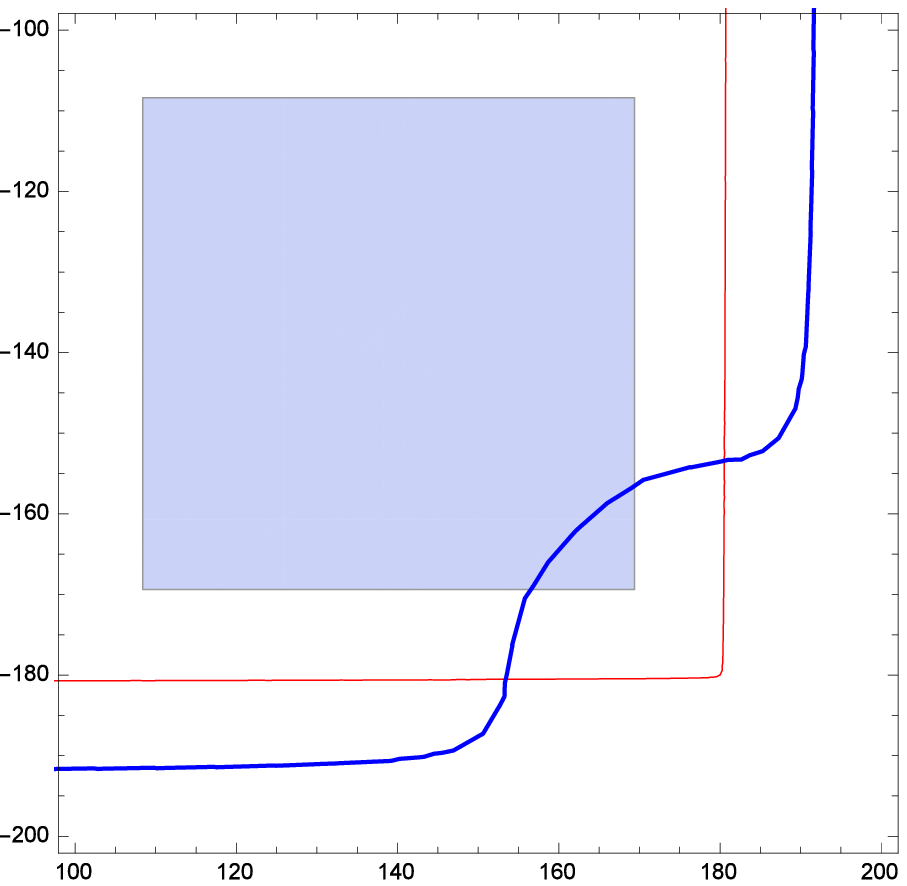,width=5.5cm}&
\psfig{figure=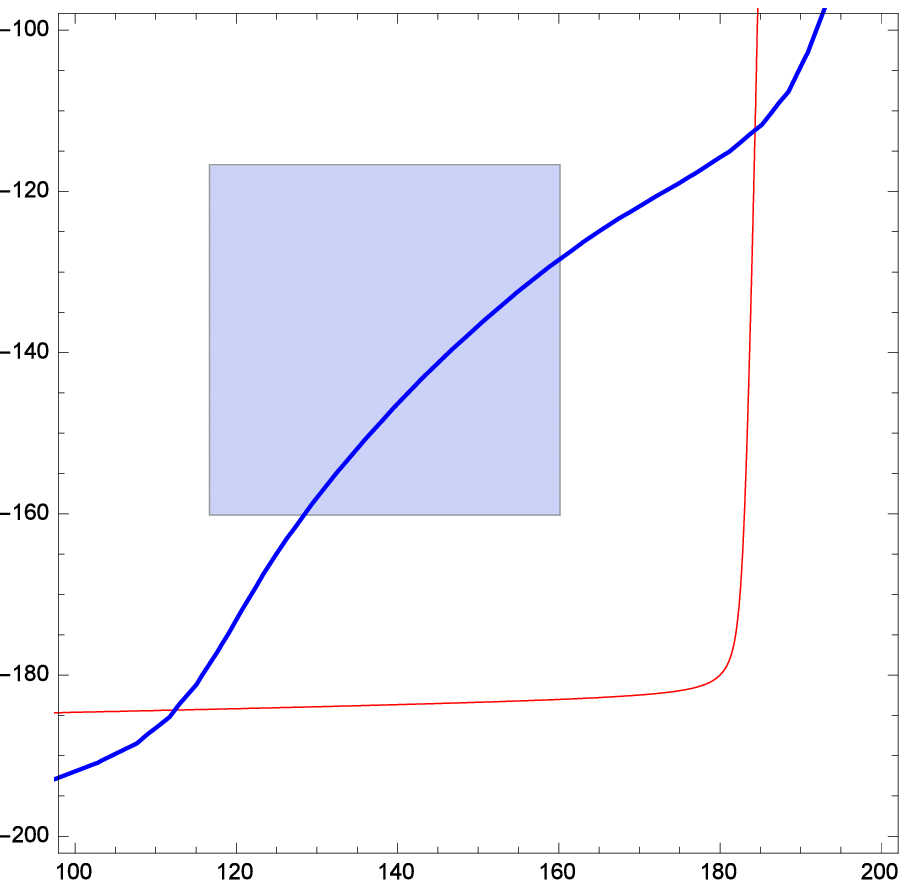,width=5.5cm}\\
(a) & (b)
\end{tabular}
\end{center}
\caption{The stability region for nodoids with a) $B=1.05$
and b) $B=1.25$ between equal spheres with $a=1.2$.
Blue shaded area determines the stability region w.r.t.
axisymmetric perturbations.
Solid curves represent fixed CL stability boundary
${\mathcal C}^{(n)}$ for $n=0$ (red)
and $n=1$ (blue). The boundary ${\mathcal B}^{(1)}$ 
lies outside of the shown regions.}
\label{f03}
\end{figure}

Figure \ref{f03} shows
the stability regions for convex nodoid ($S_H=1$) between two equal
solid spheres which demonstrates that only ${\mathcal C}^{(1)}$
but not ${\mathcal B}^{(1)}$ crosses the
axisymmetric stability region ${\sf Stab}_0$.

It is important to underline that asymmetric perturbations
just reduce the stability region for the nodoids but not
completely forbid their stability contrary to the statement
in \cite{Vogel2006} that
"\ldots a convex unduloidal bridge
between two balls is a constrained local energy minimum for the capillary
problem, and a convex nodoidal bridge between two balls is unstable".

\section{Discussion}\label{s7}
In this manuscript we consider an extension of the
analysis of axisymmetric menisci stability presented in
\cite{FelRub15} to the case of asymmetric perturbations.
The method itself is a development of the Weierstrass'
general method valid in case of fixed CLs \cite{Weier1927, Bolz1904}.
The asymmetric perturbations in our approach presented as an
expansion into the Fourier angular modes, the same way it was
suggested in \cite{Myshkis87}.
The stability analysis of the first perturbation mode
is made analytically for all possible setups of the solid bodies.
The case of arbitrary meniscus between two parallel plates
is considered in Section \ref{s621}, we found that
its stability coincides with ${\sf Stab}_0$.
Another significant conclusion of our computations is that
there exist stable convex nodoids between two solid spheres.

Several important facts were established using numerical solutions
of equation (\ref{h5c}) with zero BC $w(t_j)=0$ for
menisci with fixed CL, and with mixed BC $(V_j/\eta_j)w(t_j)-w'(t_j)=0$
for menisci with free CL. These are:
\begin{enumerate}
\item
The solution of Jacobi equation for
$n$-th perturbation mode with
fixed CL exists only for $B > n$.
\item
For $n > 0$ the boundary ${\mathcal C}^{(n+1)}$ lies 
outside the stability region ${\sf C}_n$, i.e., ${\sf C} = {\sf C}_0 \cap {\sf C}_1$.
\item
For $n > 0$ the boundary ${\mathcal B}^{(n+1)}$ lies 
outside the stability region ${\sf Stab}_n$, i.e.,
${\sf Stab} = {\sf Stab}_0 \cap {\sf Stab}_1$.
\end{enumerate}
Qualitatively similar result was obtained in \cite{Myshkis87} using the
analysis of the eigenvalues spectrum of the SLE for
an arbitrary perturbation mode. It would be very useful to
have a proof of the abovementioned observations.

\section*{Acknowledgements}
\label{s8}
The author is grateful to L. Fel for numerous fruitful discussions.

\appendix
\renewcommand{\theequation}{\thesection\arabic{equation}}

\section{Computation of $\chi_j$}
\label{appendix1}
\setcounter{equation}{0}
Consider a derivation of an explicit expression for
the parameter $\chi_j$ introduced in \cite{Myshkis87}
for the computation of stability region.
This quantity appears in the BC
$\chi_j w^{(n)}(t_j)+(-1)^jw'^{(n)}(t_j)=0$.
The definition of $\chi_j$ in \cite{Myshkis87} reads
\be
\chi_j \sin\theta_j = \kappa_j \cos\theta_j -\bar\kappa_j,
\label{a10}
\ee
where $\kappa_j$ and $\bar\kappa_j$ denote the
planar curvature of the meridional cross sections of the
meniscus and solid body, respectively, computed at the
$j$-th contact point $t=t_j$, where $r(t_j)=R_j(\tau_j)$.
The contact angle $\theta_j$ is determined as
$\cos\theta_j = \langle{\bf t}_j,{\bf T}_j\rangle/(|{\bf t}_j||{\bf T}_j|)$.
As for the meniscus  
it holds that $|{\bf t}_j|=1,$ we can write
\be
\cos\theta_j = (-1)^{j+1}\frac{R_j'r_j'+Z_j'z_j'}{\sqrt{R_j'^2+Z_j'^2}},
\quad
\sin\theta_j = \frac{z_j'R_j'-r_j'Z_j'}{\sqrt{R_j'^2+Z_j'^2}} =
\frac{\eta_j}{\sqrt{R_j'^2+Z_j'^2}},
\label{a11}
\ee
where the prime ${}'$ denotes differentiation w.r.t. $t$ when it acts on ${\bf r}$
and w.r.t. $\tau$ when it acts on ${\bf R}$.
The curvature $\kappa$ of the planar curve defined
parametrically $\{r(t),z(t)\}$ reads
$\kappa = (r'z''-z'r'')/(r'^2+z'^2)^{3/2},$
so that we obtain
\be
\kappa_j = r_j'z_j''-z_j'r_j'' = - r_j''/z_j',
\quad
\bar\kappa_j = (-1)^{j+1}\frac{R_j'Z_j''-Z_j'R_j''}{(R_j'^2+Z_j'^2)^{3/2}},
\label{a12}
\ee
where we use the relation $r_j'r_j''+z_j'z_j''=0$.
Substituting (\ref{a12}) into (\ref{a10}) we find
\be
\chi_j\eta_j =
(-1)^{j+1}\left[
R_j'z_j''-Z_j'r_j''-\frac{R_j'Z_j''-Z_j'R_j''}{R_j'^2+Z_j'^2}
\right] = (-1)^{j+1}V_j,
\quad
\chi_j =
(-1)^{j+1}V_j/\eta_j.
\label{a13}
\ee



\section{Stability region ${\sf C}$ for menisci with fixed CL}
\label{appendix2}
\setcounter{equation}{0}

In the case of fixed CL the solution $\bar{w}^{(n)}(t)$ of the
Jacobi equation (\ref{h5c}) with zero BC $\bar{w}^{(n)}(t_j)=0$
can be expressed as a superposition of two fundamental solutions.
When one of these two solutions,
say, $\bar{w}^{(n)}_2$ is known, the other one can be
found as $\bar{w}^{(n)}_1=U^{(n)}\bar{w}^{(n)}_2,$ where
$U'^{(n)}=g/(r [\bar{w}^{(n)}_2]^2),$
and $g$ is a constant depending on the parameter $B$ (see \cite{FelRub15}).
For example, for $n=1$ we have
$$
U^{(1)}(t) = \frac{\bar{w}^{(1)}_1(t)}{\bar{w}^{(1)}_2(t)}
= z(t) + \frac{r(t)r'(t)}{z'(t)}.
$$
Using this representation in (\ref{eq7}) we write it
as $U^{(n)}(t_1) = U^{(n)}(t_2),$ where $t_1 > t_2$.
For given $t_2$ introduce a function $\Psi^{(n)}(t)=U^{(n)}(t)-U^{(n)}(t_2),$
and write the condition on the boundary ${\mathcal C}^{(n)}$
as $\Psi^{(n)}(t^{(n)}_1)=0$. As we have $\Psi'^{(n)}=U'^{(n)},$
this derivative retains its sign but it can diverge
(when $\bar{w}^{(n)}_2=0$ or $r=0$ for
a spherical meniscus at $B=1$). The condition
$\bar{w}^{(n)}_2=0$ indicates that the function $\Psi^{(n)}(t)$
might vanish, so that a root $t^{(n)}_1$ exists.
It is easy to see that for $n=1$ the relation
$\bar{w}^{(n)}_2=0$ can be valid only for $B > 1$, so that
for unduloids the boundary ${\mathcal C}^{(1)}$ does not exist.
For $B=n>1$ there are no boundaries
${\mathcal C}^{(k)}$ with $1 \le k \le n$; it follows from the
fact that $\tilde{w}^{(n)}_2(t)$ never vanishes while
$\tilde{w}^{(n)}_1(t)$ is always positive.

For $n>1$ the solution of (\ref{h5c}) with zero BC
can be found numerically by employing the shooting method when
the above conditions are replaced by
$w^{(n)}(t_2)=0, \; w'^{(n)}(t_2)=1,$ used as initial conditions (IC)
for numerical integration of equation (\ref{h5c}).
The resulting solution is used to find a value
$t=t^{(n)}_1$ at which $w(t)$ vanishes,
and (in case such a value exists) it provides
a point $(t^{(n)}_1,t_2)$ belonging to the stability region boundary
for $n$-th perturbation mode.
The set of such points
completely defines the boundary ${\mathcal C}^{(n)}$.

The computational analysis of equation (\ref{h5c}) shows that
$t^{(n)}_1$ exists only for $B > n$ (see Appendix \ref{appendix0}). It is instructive for
given value of $t_2$
compare the values $t^{(n)}_1$ and $t^{(n+1)}_1$.
It appears that it holds always that $t^{(n+1)}_1 > t^{(n)}_1,$
which implies that the boundary ${\mathcal C}^{(n+1)}$
lies outside of the region ${\sf C}_n$ bounded by
${\mathcal C}^{(n)}$. This observation indicates that
the stability region ${\sf C}$ for menisci with fixed CL
is determined exclusively by intersection
${\sf C} = {\sf C}_0 \cap {\sf C}_1$ of the regions
for axisymmetric and first asymmetric modes.
This result confirms the statement made in
\cite{Myshkis87} about the stability region for
the case of fixed CL.

\section{Stability region ${\sf Stab}$ for menisci with free CL}
\label{appendix3}
\setcounter{equation}{0}

The relation (\ref{p02}) which determines the stability boundaries
${\mathcal B}^{(n)}$ employs matrices $A_k^{(n)}$ that depend
on the fundamental solutions $w^{(n)}_i(t)$ and their derivatives.
Using the representation $w^{(n)}_1(t)=U^{(n)}(t)w^{(n)}_2(t),$
we rewrite (\ref{p02}) for $n>0$ as
$$
(U_1-U_2)[V_1V_2-\eta_1G_1V_2+\eta_2G_2V_1+\eta_1\eta_2G_1G_2]
-\eta_1V_2U'_1+\eta_2V_1U'_2+
\eta_1\eta_2[G_2U'_1-G_1U'_2]=0,
$$
where
$$
U_j=U^{(n)}(t_j),\quad
U'_j=U'^{(n)}(t_j),\quad
G_j=w'^{(n)}_2(t_j)/w^{(n)}_2(t_j).
$$
The above relation can be rewritten as
$$
(U_1-U_2)(V_1-\eta_1G_1)(V_2-\eta_2G_2)
-\eta_1U'_1(V_2-\eta_2G_2)
-\eta_2U'_2(V_1-\eta_1G_1)=0,
$$
leading to the condition
\be
\Phi_1 = \Phi_2,
\quad
\Phi_j = U_j - \frac{\eta_jU'_j}{V_j-\eta_jG_j}.
\label{a3e1}
\ee
Returning to the original notation for the fundamental solutions
we find a compact expression for (\ref{p02}) in the form
\be
\Phi^{(n)}(t_1)=\Phi^{(n)}(t_2),
\quad
\Phi^{(n)}(t) =
\frac{Vw^{(n)}_1-\eta w'^{(n)}_1}{Vw^{(n)}_2-\eta w'^{(n)}_2}.
\label{a3e2}
\ee
It is easy to see that the condition (\ref{a3e2}) is equivalent to
(\ref{s44e4}) as expected.
From the computational perspective the problem of finding a point
$(t_1,t_2)$ belonging to the boundary ${\mathcal B}^{(n)}$ is reduced to
a problem of finding the first zero $t^{(n)}_1 > t_2$ of the function
$\Psi^{(n)}(t)=\Phi^{(n)}(t)-\Phi^{(n)}(t_2)$.
Setting in (\ref{a3e2}) $\eta=0$ we obtain $\Phi^{(n)}(t)=U^{(n)}(t),$
and we recover the condition for the
stability boundary ${\mathcal C}^{(n)}$
derived in Appendix \ref{appendix2} for the menisci with fixed CL.

The numerical computations show that the stability boundary
${\mathcal B}^{(1)}$ might exist for $B<1$
but it appears that it does not intersect ${\sf Stab}_0$.
This observation implies that asymmetric perturbations with free CL
{\it do not} affect unduloid stability region ${\sf Stab}_0$
constructed using the analysis
of axisymmetric perturbations only. In other words, for all unduloids
we have ${\sf Stab}={\sf Stab}_0$, because any asymmetric perturbation
is less dangerous than axisymmetric one.
In case of nodoids with $B>1$ we found that ${\mathcal B}^{(1)}$
also does not intersect ${\sf Stab}_0$, so that only
${\mathcal C}^{(1)}$ might lead to reduction of the stability region.

\section{Analysis of Jacobi equation}
\label{appendix0}
\setcounter{equation}{0}

Consider homogeneous Jacobi equation (\ref{k05}) and use a
replacement $w=y/r$ to produce
\be
r^2 y'' - r r' y' +(B^2-n^2+rz')y = 0.
\label{a01}
\ee
Substituting an ansatz $y = a_0 + a_1 \cos t + a_2 \sin t,$
into (\ref{a01}) we arrive at
$$
[a_0(1+B^2-n^2)-a_1 B] -(a_0 B - a_1 n^2)\cos t -a_2 n^2 \sin t =0,
$$
which leads to a system
\be
a_0(1+B^2-n^2)-a_1 B = 0,
\quad
a_0 B - a_1 n^2=0,
\quad
a_2 n^2 = 0.
\label{a02}
\ee
Direct substitution shows that
for $n=0$ we have $a_0=a_1=0,$ and we reproduce the solution (\ref{h5d}).
With $n=1$ we find $a_0=1,\; a_1=B\; a_2=0,$ and we arrive at (\ref{h5e}).
Finally, setting $B=n,$ we obtain $a_0=B,\; a_1=1\; a_2=0,$
and generate the solution (\ref{h5e1}).

The IC $w_1(0)=0,\; w'_1(0)=const>0,$ for (\ref{k05}) convert into
$y_1(0)=0,\; y'_1(0)=const>0,$ while
the IC $w'_2(0)=0,\; w_2(0)=const>0,$ lead to
$y'_2(0)=0,\; y_2(0)=const>0$.
We performed numerical integration
and found that for given value of $n$ the solutions
to (\ref{a01}) have qualitatively different behavior
in two regions -- $B < n,$ and $B > n$. These
solutions are separated by the solution (\ref{h5e1}).

First, we found
that for $B < n,$ both $y_1(t)$ and $y_2(t)$ are
positive functions and for $t \gg 1$ it holds
asymptotically that $y_1(t)\sim c(B,n)y_2(t),$ where positive constant $c$
depends on both $B$ and $n$. This observation implies that
the function $\Psi^{(n)}$ introduced in Appendix \ref{appendix3}
tends to constant for large $t$, and, moreover, we
observe $\Psi^{(n)}\approx U^{(n)}$. This leads to a conclusion
that ${\mathcal C}^{(n)}$ 
does not exist for $B < n,$ so that the stability region with fixed CL
is found as
${\sf C} = \cap_{k=0}^{n-1} {\sf C}_k$.

\begin{figure}[h!]
\begin{center}
\begin{tabular}{cc}
\psfig{figure=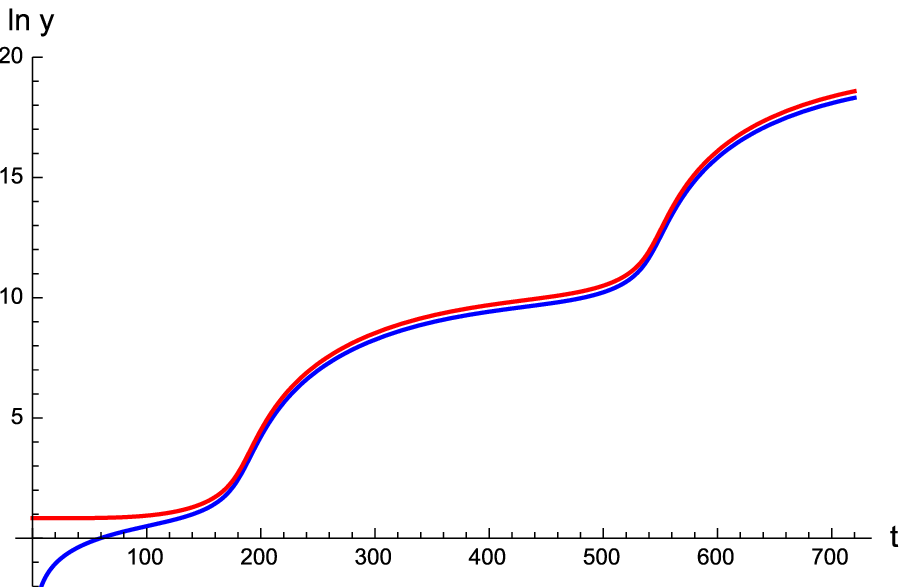,width=6.0cm}&
\psfig{figure=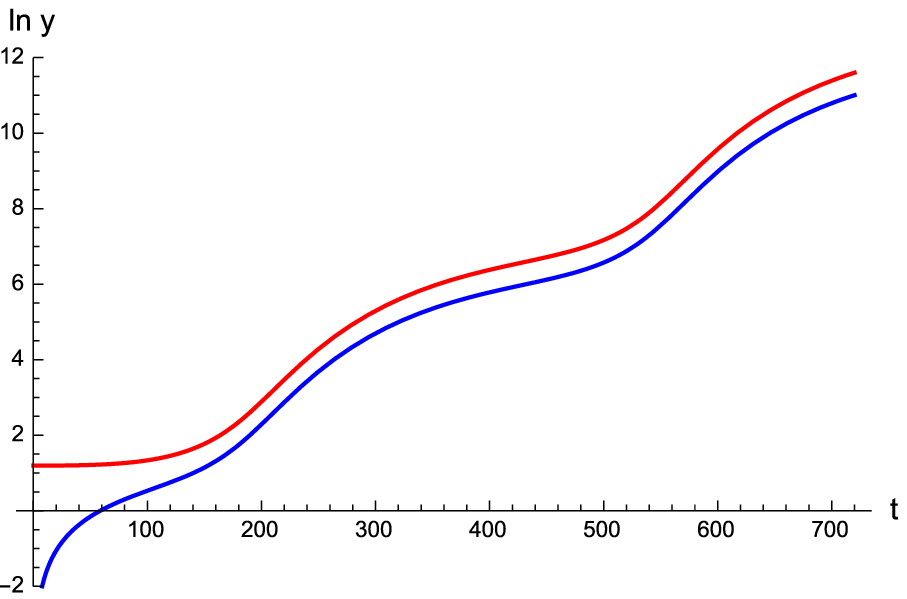,width=6.0cm}\\
(a) & (b)
\end{tabular}
\end{center}
\caption{The logarithm of the fundamental
solutions $\ln y_1(t)$ (blue) and $\ln y_2(t)$ (red)
of the Jacobi homogeneous equation (\ref{k05})
for (a) $n=2,\; B=1.3,$ and (b) $n=3,\; B=2.3$.
}
\label{f04}
\end{figure}

In the other case $B > n,$
we observed that both $\bar{w}^{(n)}_i(t)$ change sign,
so that the function $U^{(n)}$ changes sign too and
thus the curve ${\mathcal C}^{(n)}$ exists.
Similarly, the function $\Psi^{(n)}$ changes sign and its first zero
determines the curve ${\mathcal B}^{(n)}$.
The numerical simulations showed that the first root of the
function $U^{(n)}$ can be approximated by
$t^{(n)}_1 \approx a(n)/\sqrt{\epsilon},$ where $0 < \epsilon=B-n \le 1,$
and $a(n+1) > a(n)$. A similar dependence of
$t^{(n)}_1-t_2 \approx a(n)/\sqrt{\epsilon}$ is valid
for nonzero $t_2$.
This implies that
$t^{(n+1)}_1-t_2 > t^{(n)}_1-t_2$ for all $n>0$,
and the boundary ${\mathcal C}^{(n+1)}$
lies outside of the region ${\sf C}_n$ bounded by
${\mathcal C}^{(n)}$.



\end{document}